\documentclass[12pt]{elsarticle}                                          
\usepackage{axodraw4j}
\usepackage{hyperref}
\usepackage{graphicx}                                                         
\usepackage{a41}                                                                
\usepackage{xcolor}                     
\usepackage[rflt]{floatflt}
\usepackage{float}
\usepackage{lscape}
\usepackage{hyperref}
\hypersetup{colorlinks=true, 
linkcolor=blue, filecolor=blue, urlcolor=mygreen}
\usepackage{breakurl} 
\usepackage{times}     %
\usepackage{array}
\usepackage[english]{babel}
\usepackage[latin1]{inputenc}
\usepackage[T1]{fontenc}
\usepackage{ae}
\usepackage{url}
\usepackage{amsmath, amsthm, amssymb}
\usepackage{slashed}

\usepackage{rotating}
\usepackage{graphicx}
\usepackage{comment}
\newcounter{mmacnt}
\def\restartmma{\setcounter{mmacnt}{0}}
\restartmma \catcode`|=\active
\def|#1|{\mathrm{#1}}
\catcode`|=12
\newenvironment{mma}{
\par\smallskip
\catcode`|=\active
\parskip=0pt\parindent=0pt 
\small
\def\In##1\\{%
\def\linebreak{\hfill\break\null\qquad}%
\refstepcounter{mmacnt}
\hangindent=2.5em\hangafter=0
\leavevmode
\llap{\tiny\sffamily In[\arabic{mmacnt}]:=\kern.5em}%
\mathversion{bold}\footnotesize$
\displaystyle##1$\normalsize
\mathversion{normal}\par
 }%
\def\Print##1\\{%
\def\linebreak{\hfill\break}%
\hangindent=2.5em\hangafter=0
\leavevmode ##1\par}%
\def\Out##1\\{%
\def\linebreak{$\hfill\break\null\hfill$}%
\kern\abovedisplayskip\par
\hangindent=2.5em\hangafter=0
\leavevmode
\llap{\tiny\sffamily Out[\arabic{mmacnt}]=\kern.5em}
\footnotesize$\displaystyle##1$
\normalsize\hfill\null\par
\kern\belowdisplayskip
}%
\def\Warning##1##2\\{%
\def\linebreak{\hfill\break}%
\hangindent=2.5em\hangafter=0
\leavevmode
{\scriptsize##1 : ##2}\par}%
}{%
\par\smallskip
}

\usepackage{color}
\newenvironment{fshaded}{%
\MakeFramed {\FrameRestore}
}%
{\endMakeFramed}

\makeatletter
\def\ps@pprintTitle{%
\let\@oddhead\@empty
\let\@evenhead\@empty
\def\@oddfoot{\reset@font\hfil\thepage\hfil}
\let\@evenfoot\@oddfoot
}
\makeatother
\usepackage{tikz}
\usetikzlibrary{matrix}
\allowdisplaybreaks[4]

\begin{document}
\begin{frontmatter}
\title{\Large
\textbf{Processes $\gamma \gamma
\rightarrow \phi_i\phi_j$
in Inert Higgs Doublet Models and 
Two Higgs Doublet Models}}
\author[1,2]{Khiem Hong Phan}
\ead{phanhongkhiem@duytan.edu.vn}
\author[1,2]{Dzung Tri Tran}
\author[3]{Thanh Huy Nguyen}
\address[1]{\it Institute of Fundamental
and Applied Sciences, Duy Tan University,
Ho Chi Minh City $70000$, Vietnam}
\address[2]{Faculty of Natural Sciences,
Duy Tan University, Da Nang City $50000$,
Vietnam}
\address[3]
{\it VNUHCM-University of Science, 
$227$ Nguyen Van Cu, District $5$, 
Ho Chi Minh City $70000$, Vietnam}
\pagestyle{myheadings}
\markright{}
\begin{abstract} 
In this paper, we present a phenomenological analysis of one-loop induced processes $\gamma \gamma \rightarrow \phi_i\phi_j$, where the CP-even Higgs bosons are denoted as $\phi_{i,j} \equiv h,~H$, in high-energy photon-photon collisions, within the frameworks of the Inert Higgs Doublet Model and the Two Higgs Doublet Model. The total cross sections are evaluated as functions of the center-of-mass energy, finding that the cross sections for the considered processes in all the models under investigation are significantly enhanced around the threshold for charged Higgs boson pair production ($\sim 2 M_{H^\pm}$). Furthermore, the enhancement factors, defined as the ratios of cross sections of $\gamma \gamma \rightarrow \phi_i\phi_j$ in the investigated models to those for $\gamma \gamma \rightarrow hh$ in the Standard Model, are examined in the relevant regions of the model's parameter space. In the Inert Higgs Doublet Model, the factors are studied in the parameter space of $(M_{H^\pm},~\mu^2_2)$ and $(M_{H^\pm},~\lambda_2)$. In the Two Higgs Doublet Model, the factors are examined in the planes defined by $(M_{H^\pm},~t_{\beta})$ as well as in the space of the charged Higgs mass $M_{H^\pm}$ and the soft-breaking $Z_2$ parameter $m_{12}^2$. Two scenarios characterized by $c_{\beta-\alpha} > 0$ and $c_{\beta-\alpha} < 0$ are studied in further detail. The factors exhibit distinct behaviors between these two scenarios. As a result, it is possible to discriminate between them at future colliders. The dependence of the cross section for the process $\gamma \gamma \rightarrow hH$ on $m_{12}^2$ provides a potential probe of the soft $Z_2$-breaking scale in the Two Higgs Doublet Model.
\end{abstract}
\end{frontmatter}
\section{Introduction}
Measuring scalar Higgs self-couplings, including Standard Model-like Higgs trilinear and quadratic couplings, as well as the couplings between scalar Higgses in many beyond-the-Standard-Model (BSM) scenarios, plays a key role in determining the scalar potential. We can subsequently answer the question of the origin of electroweak symmetry breaking (EWSB). In this scheme, Higgs boson pair production and multi-scalar Higgs production should be measured precisely at future colliders.
Recently, searches for Higgs boson pair production in final states such as two bottom quarks associated with two photons, four bottom quarks, etc., in proton-proton collisions have been performed at the Large Hadron Collider (LHC), as in~\cite{ATLAS:2021ifb, ATLAS:2014pjm, ATLAS:2015zug, ATLAS:2015sxd, CMS:2017rpp, ATLAS:2018dpp, CMS:2018tla, CMS:2020tkr, CMS:2022cpr, ATLAS:2022xzm, ATLAS:2023qzf, ATLAS:2024ish}. It is well known that measurements of Higgs self-couplings are rather challenging at the LHC. The results from the study in~\cite{Baur:2002rb, Baur:2002qd} show that the expected accuracy in the measurement of trilinear Higgs self-couplings would be about $20\%$\textendash$30\%$ at the high luminosity of $3000$ fb$^{-1}$.
We know that the physics program of future lepton colliders (LC) will be complementary to that of the LHC in many aspects, as studied in~\cite{LHCLCStudyGroup:2004iyd}. Furthermore, the LC can significantly improve LHC measurements in many cases.
More importantly, photon-photon collisions are an option at the LC~\cite{ILC:2013jhg, Shiltsev:2019rfl}, where scalar Higgs pair production ($\phi_i\phi_j$) can be measured via the channels $f\bar{f} \rightarrow f\bar{f} \gamma^*\gamma^* \rightarrow f\bar{f} \phi_i\phi_j$ for $f \equiv e, \mu$. In this regard, the LC could open a window for probing new physics signals through multi-scalar Higgs production.

From a theoretical calculation perspective,
one-loop corrections to double Higgs production at the LHC have been calculated in the Standard Model (SM), the Higgs Extensions of the Standard Models (HESM), as well as in other BSM frameworks by many research groups. For example, it is worth mentioning
notable works in~\cite{Arhrib:2009hc, Grigo:2013rya, Shao:2013bz, Ellwanger:2013ova, Han:2013sga, Barr:2013tda, deFlorian:2013jea, Haba:2013xla, Cao:2014kya, Enkhbat:2013oba, Li:2013flc, Frederix:2014hta, Baglio:2014nea, FerreiradeLima:2014qkf, Hespel:2014sla, Barger:2014taa, Grigo:2014jma, Maltoni:2014eza, Goertz:2014qta, Azatov:2015oxa, Papaefstathiou:2015iba, Grober:2015cwa, deFlorian:2015moa, He:2015spf, Grigo:2015dia, Zhang:2015mnh, Agostini:2016vze, Grober:2016wmf, Degrassi:2016vss, Kanemura:2016tan, deFlorian:2016uhr, Borowka:2016ypz, Bishara:2016kjn, Cao:2016zob, Nakamura:2017irk, Grober:2017gut, Heinrich:2017kxx, Jones:2017giv, Davies:2018ood, Goncalves:2018qas, Chang:2018uwu, Buchalla:2018yce, Bonciani:2018omm, Banerjee:2018lfq, AH:2018tzg, Baglio:2018lrj, Davies:2019xzc, Davies:2019dfy, Chen:2019lzz, Chen:2019fhs, Baglio:2020ini, Wang:2020nnr, Abouabid:2021yvw, Davies:2022ram, He:2022lhc, AH:2022elh, Iguro:2022fel, Alioli:2022dkj, Davies:2023npk, Bagnaschi:2023rbx, Davies:2024znp}, for further reviews in~\cite{Brigljevic:2024vuv} and the related references therein.
Additionally, one-loop corrections for $hh$ production in high-energy $\gamma\gamma$ collisions in the SM, HESM, and other BSM models have been computed in Refs.~\cite{Jikia:1992mt, Sun:1995rd, Zhu:1997nz, Zhu:1998rh, Gounaris:2000ja, Zhou:2003ss, Cornet:2008nq, Asakawa:2008se, Asakawa:2009ux, Takahashi:2009sf, Hodgkinson:2009uj, Arhrib:2009gg, Asakawa:2010xj, Hernandez-Sanchez:2011idv, Ma:2011zzb}.
Other computations for one-loop corrections to Higgs boson pair production at future linear lepton colliders, including multi-TeV muon colliders, have been performed in Refs.~\cite{Sola:2011ex, Heng:2013wia, Chiesa:2021qpr, Samarakoon:2023crt} and the additional references therein.
Furthermore, double pseudo-scalar Higgs $A^0A^0$ production at a $\gamma\gamma$ collider in the Two Higgs Doublet Model has been reported in Ref.~\cite{Demirci:2019kop}.
One-loop contributions to the process $\gamma \gamma \rightarrow \phi_i\phi_j$ with CP-even Higgses $\phi_{i,j} \equiv h,~H$ in high-energy $\gamma\gamma$ collisions, valid for the Higgs Extensions of the Standard Models (HESM), are necessary. In our previous paper, analytic formulas for the concerned processes were reported~\cite{Phan:2024vxm}.

In this paper, we present the phenomenological analysis of these processes at high-energy $\gamma\gamma$ collisions in the Inert Higgs Doublet Model (IHDM) and the Two Higgs Doublet Model (THDM).
In the phenomenological analysis, cross sections are shown as functions of center-of-mass energies.
Furthermore, the enhancement factors, defined as the ratios of cross sections of $\gamma \gamma \rightarrow \phi_i\phi_j$ in the HESMs to the corresponding ones for $\gamma \gamma \rightarrow hh$ in the Standard Model, are examined in the parameter space of the models under consideration.
The paper is structured as follows. We briefly review the HESM in the next section. The phenomenological studies for the HESM are discussed in section 3. In section 4, we present the conclusion and outlook of the paper. In appendices A and B, we derive the couplings that appear in the calculations.
\section{Higgs Extensions of         
the Standard Model}                  
Two typical Higgs extensions of
the Standard Model are studied in this paper.
The first model is the Inert Higgs Doublet Model,
which is reviewed in subsection 2.1.
We then discuss the Two Higgs Doublet Model in
subsection 2.2.
\subsection{Inert Higgs Doublet Models}
In the IHDM, an inert scalar $SU(2)_L$ doublet
is included in the potential of the SM.
The inert scalar particles
serve as dark matter candidates.
For a review of the theory
and phenomenological studies
of the IHDM in detail,
we refer to the following
papers~\cite{Borah:2012pu,
Gustafsson:2012aj,Arhrib:2012ia,
Klasen:2013btp,Krawczyk:2013jta,
Arhrib:2014pva,Chakrabarty:2015yia,
Ilnicka:2015jba,Datta:2016nfz,
Kalinowski:2018ylg,Dercks:2018wch,
Chiang:2012qz,Benbrik:2022bol}.
The scalar potential of the IHDM
is expressed as follows:
\begin{eqnarray}
\label{VIHDM}
\mathcal{V}_{\textrm{IHDM}}
(\phi_1,\phi_2)
&=& \mu_1^2 |\phi_1|^2
+ \mu_2^2 |\phi_2|^2
+ \lambda_1 |\phi_1|^4
+ \lambda_2 |\phi_2|^4
+  \lambda_3 |\phi_1|^2
|\phi_2|^2
+ \lambda_4 |\phi_1^\dagger
\phi_2|^2
\nonumber \\
&&
+ \frac{\lambda_5}{2}
\left\{ (\phi_1^\dagger \phi_2)^2
+ {\rm h.c} \right\}.
\end{eqnarray}
The potential is conserved with respect to the so-called global $Z_2$-symmetry, e.g., $\phi_1 \leftrightarrow +\phi_1$, $\phi_2 \leftrightarrow -\phi_2$. In this case, the scalar $\phi_2$ is odd, and $\phi_1$ and all particles in the SM are even under the $Z_2$-transformation. As mentioned, the $Z_2$-symmetry remains unbroken after the EWSB. The field $\phi_2$ then has a zero vacuum expectation value (VEV), while the field $\phi_1$ develops a non-zero VEV ($v$). The two scalar fields are then expanded around their VEVs for the EWSB as follows:
\begin{eqnarray}
\phi_1 &=&
\left (\begin{array}{c}
G^\pm \\
\frac{1}{\sqrt{2}}(v + h + i G^0) \\
\end{array} \right),
\\
\phi_2 &=&
\left( \begin{array}{c}
H^\pm\\
\frac{1}{\sqrt{2}}(H + i A^0) \\
\end{array} \right).
\end{eqnarray}
Where the VEV is fixed at $v \sim 246$ GeV (as in the SM case). The Goldstone bosons $G^{\pm},\; G^0$ give the masses for the $W^{\pm}$ boson and the $Z$ boson, respectively. There is no mixing between $h$ and $H$. The physical spectrum of the IHDM after the EWSB consists of three neutral scalar physical states: two CP-even Higgses $h,~H$ and a CP-odd Higgs $A_0$. Furthermore, we have a pair of singly charged Higgs bosons $H^\pm$ in this model. In the spectrum, the neutral scalar Higgs $h$ is identical to the SM-like Higgs boson discovered at the LHC. The masses of the scalar bosons are calculated from the model parameters as follows:
\begin{eqnarray}
\label{scalarIHDM}
M_h^2 &=& - 2 \mu_1^2 = 2 \lambda_1 v^2,
\\
M_{H}^2 &=& \mu_2^2 + \frac{v^2}{2} \lambda_L,
\\
M_{A^0}^2 &=& \mu_2^2 + \frac{v^2}{2} \lambda_R,
\\
M_{H^{\pm}}^2 &=& \mu_2^2 + \frac{v^2}{2} \lambda_3.
\end{eqnarray}
We have used $\lambda_{L/R} = \lambda_3 + \lambda_4 \pm \lambda_5$. As mentioned in the previous paragraphs, the global $Z_2$-symmetry is unbroken after the EWSB. The "inert" Higgs bosons such as $H^{\pm}$, $A^0$, and $H$ have an odd number under the $Z_2$-transformation. Subsequently, the "inert" Higgs bosons do not couple to the SM particles. Therefore, the lightest neutral scalar bosons may be considered as dark matter candidates.

As a consequence of the unbroken $Z_2$-symmetry,
the Yukawa Lagrangian of the IHDM must be the same as
that of the SM. In detail, the Yukawa Lagrangian
is expressed as follows:
\begin{eqnarray}
\label{YukawaIHDM}
\mathcal{L}^{\textrm{IHDM}}_{\rm Yukawa}
= -\sum_{f=u,d,\ell}
g_{hff}\cdot h \bar{f} f
+ \cdots,
\end{eqnarray}
where the Yukawa coupling is given by $g_{hff} = m_f/v$ for fermion $f$. All the couplings involving  the computed processes $\gamma \gamma \rightarrow \phi_i \phi_j$ in the IHDM are listed in Table~\ref{IHDM-coupling} (for all physical couplings) and Table~\ref{IHDM-couplingGpm} (for unphysical particles). We emphasize that the processes $\gamma \gamma \rightarrow hH$ are forbidden in the IHDM due to the $Z_2$-symmetry. Therefore, we have only $\phi_i \phi_j \equiv hh,~HH$ in this case. The detailed expressions for all the concerned couplings are derived in appendix A.
\begin{table}[H]
\centering
\begin{tabular}
{
l@{\hspace{1.5cm}}
l@{\hspace{1.5cm}}
l }
\hline\hline
Vertices 
& Notations 
& Coupling
\\
\hline\hline
$hW_{\mu}W_{\nu}$
& $g_{hWW} 
\cdot g_{\mu\nu}$
& $i
\left(
\dfrac{2M_W^2}{v}
\right)
\cdot g_{\mu\nu}
$
\\
\hline
$hZ_{\mu}Z_{\nu}$
&
$g_{hZZ}
\cdot g_{\mu\nu}$
&$i
\left(
\dfrac{M_Z^2}{v}
\right)
\cdot g_{\mu\nu}
$
\\ \hline
$hH^\pm{H^\mp}$
&
$g_{hH^{\pm}H^{\mp}}$
&
$i
\dfrac{2(\mu_2^2-M_{H^\pm}^2)}{v}
$
\\
\hline
$Z_{\mu}
H^{\pm}(p_1)
H^{\mp}(p_2)$
&
$g_{ZH^{\pm}H^{\mp}} \cdot
(p_{1}-p_{2})_{\mu}$
&
$i
\left(
\dfrac{M_Zc_{2W}}{v}
\right)
\cdot
(p_{1}-p_{2})_{\mu}
$
\\ \hline
$
A_{\mu}
H^{\pm}(p_1)
H^{\mp}(p_2)
$
&
$g_{AH^{\pm}H^{\mp}}
\cdot
(p_{1}-p_{2})_{\mu}
$
&
$i
\left(
\dfrac{M_Zs_{2W}}{v}
\right)
\cdot
(p_{1}-p_{2})_{\mu}
$
\\ \hline
$hhh$
&
$g_{hhh}$
& $
-i
\left(
\dfrac{3M_h^2}{v}
\right)
$
\\
\hline
$hHH$
&
$g_{hHH}$
&
$i
\dfrac{2(\mu_2^2-M_H^2)}{v}
$
\\
\hline
$H(p_2)H^{\pm}(p_1)W^{\mp}_{\mu}$
&
$g_{HH^{\pm}W}
\cdot (p_1-p_2)_\mu$
&
$
\pm i
\left(
\dfrac{M_W}{v}
\right)
\cdot
(p_1-p_2)_\mu
$
\\
\hline
$H^{\pm}H^{\mp}
A_{\mu}A_{\nu}$
&
$g_{AAH^{\pm}H^{\mp}}
\cdot
g_{\mu\nu}$
&
$i
\left(
\dfrac{M_Z^2s_{2W}^2}{v^2}
\right)
\cdot
g_{\mu\nu}
$
\\
\hline
$HH^{\pm}
W^{\mp}_{\nu}A_{\mu}$
&
$g_{HH^{\pm}WA}
\cdot
g_{\mu\nu}$
&
$i
\left(
\dfrac{2M_Z^2c_W^2s_W}{v^2}
\right)
\cdot
g_{\mu\nu}
$
\\
\hline
$hhH^{\pm}H^{\mp}$
&
$g_{hhH^{\pm}H^{\mp}}$
&
$i
\dfrac{2(\mu_2^2-M_{H^\pm}^2)}
{v^2}
$
\\
\hline
$HHH^{\pm}H^{\mp}$
&
$g_{HHH^{\pm}H^{\mp}}$
&
$-2i\lambda_2$
\\
\hline
$hhW^{\pm}_{\mu}W^{\mp}_{\nu}$
&
$g_{hhWW}\cdot g_{\mu\nu}$
&
$i
\left(
\dfrac{2M_Z^2c_W^2}{v^2}
\right)
\cdot g_{\mu\nu}
$
\\
\hline
$HHW^{\pm}_{\mu}W^{\mp}_{\nu}$
&
$g_{HHWW}\cdot
g_{\mu\nu}$
&
$i
\left(
\dfrac{2M_Z^2c_W^2}{v^2}
\right)
\cdot
g_{\mu\nu}
$
\\
\hline
\hline
\end{tabular}
\caption{
\label{IHDM-coupling}
We list all couplings 
(physical couplings)
contributing to 
$ \gamma \gamma\rightarrow 
\phi_i\phi_j$
for the IHDM. 
}
\end{table}

\begin{table}[H]
\centering
\begin{tabular}{
l@{\hspace{1.5cm}}
l@{\hspace{1.5cm}}
l}
\hline\hline
Vertices 
& Notations 
& Coupling
\\
\hline\hline
$A_{\mu}W^{\pm}_{\nu}G^{\mp}$
&
$g_{AW^{\pm}G^{\mp}}
\cdot g_{\mu\nu}$
&
$i
\left(
\dfrac{2M_Z^2c_W^2s_W}{v}
\right)
\cdot g_{\mu\nu}
$
\\
\hline
$HH^{\pm}G^{\mp}$
& $g_{HH^{\pm}G^{\mp}}$
& $i
\dfrac{M_{H^\pm}^2-M_H^2}{v}$
\\
\hline
$hhG^{\pm}G^{\mp}$
&
$g_{hhG^{\pm}G^{\mp}}$
&
$
-i
\left(
\dfrac{M_h^2}{v^2}
\right)
$
\\
\hline
$HHG^{\pm}G^{\mp}$
&
$g_{HHG^{\pm}G^{\mp}}$
&
$i
\dfrac{2(\mu_2^2
-M_{H^\pm}^2)}{v^2}
$
\\
\hline
$A_{\mu}A_{\nu}
G^{\pm}G^{\mp}$
&
$g_{AAG^{\pm}G^{\mp}}
\cdot g_{\mu\nu}$
&
$i
\left(
\dfrac{M_Z^2s_{2W}^2}{v^2}
\right)
\cdot g_{\mu\nu}
$
\\
\hline
$A_{\mu}
G^{\pm}(p_1)
G^{\mp}(p_2)
$
&
$g_{AG^{\pm}G^{\mp}}
\cdot
(p_1-p_2)_{\mu}
$
&
$
i
\left(
\dfrac{M_Zs_{2W}}{v}
\right)
\cdot
(p_1-p_2)_{\mu}$
\\
\hline\hline
\end{tabular}
\caption{
\label{IHDM-couplingGpm}
We list all vertices 
(unphysical couplings)
in processes
$\gamma \gamma\rightarrow \phi_i\phi_j$
in the IHDM.
}
\end{table}

The parameter space of 
the IHDM for our analysis is included as
follows:
\begin{eqnarray}
\label{eq_PIHDM}
 \mathcal{P}_{\rm IHDM} =
\{\mu_2^2, \lambda_2^2, M_h^2 \sim 125.
\textrm{GeV},
M_H^2, M_{A^0}^2, M^2_{H^{\pm}}\}.
\end{eqnarray}
We are going to review the current constraints on the physical parameter space in the IHDM given in Eq.~\ref{eq_PIHDM}. The constraints for the physical parameter space can be obtained by including the theoretical conditions and the experimental data. In the perspective of the experimental data, we take into account the Electroweak Precision Tests (EWPT) of the IHDM, dark matter search at the LHC, as well as the LEP data. The topics have been studied in Refs.~\cite{Borah:2012pu, Gustafsson:2012aj, Arhrib:2012ia, Klasen:2013btp}. Additionally, the implications for loop-induced decays of the SM-like Higgs ($h$) to $V\gamma$ with $V \equiv Z, \gamma$ in the IHDM framework, e.g. the decay process $h \rightarrow \gamma \gamma$, have been reported in~\cite{Krawczyk:2013jta, Chiang:2012qz, Benbrik:2022bol}, and decay channels $h \rightarrow Z \gamma$ have been examined in~\cite{Chiang:2012qz, Benbrik:2022bol}. Furthermore, searching for signals of the IHDM at future colliders has been performed in Refs.~\cite{Arhrib:2014pva, Chakrabarty:2015yia, Ilnicka:2015jba, Datta:2016nfz, Kalinowski:2018ylg, Dercks:2018wch}. In the theoretical side, the most important theoretical constraints are obtained from the conditions that the models follow tree-level unitarity, vacuum stability, and the perturbative regime. The theoretical constraints give the limitations on the Higgs self-couplings $\lambda_i$ for $i=1,2,\cdots,5$ and $\mu_2$. Taking the theoretical and experimental constraints from the above-mentioned papers, one can select the parameter space for the IHDM as follows: we can take $5$ GeV $\leq M_H \leq 150$ GeV, $70$ GeV $\leq M_{H^{\pm}}, M_{A^0} \leq 1000$ GeV, $|\mu_2| \leq 500$ GeV, and $0 \leq \lambda_2 \leq 8 \pi$.
\subsection{Two Higgs Doublet Models} 
We now considerthe second kind of Higgs sector extension of the Standard Model, namely the Two Higgs Doublet Models, in this paper. For a comprehensive review of the theory and phenomenological studies of the THDM, we refer the reader to Ref.~\cite{Branco:2011iw} for further information. The model is briefly summarized
in this section. A complex Higgs doublet with hypercharge
$Y = 1/2$ is added to the scalar sector of the SM. The scalar potential takes the form:
\begin{eqnarray}
\label{V2HDM}
\mathcal{V}_{\textrm{THDM}}(\phi_1,\phi_2) &=&
m_{11}^2\phi_1^\dagger \phi_1
+ m_{22}^2\phi_2^\dagger \phi_2-
\Big[
m_{12}^2\phi_1^\dagger \phi_2
+{\rm h.c.}
\Big]
+ \frac{\lambda_1}{2}
(\phi_1^\dagger \phi_1)^2
+\frac{\lambda_2}{2}
(\phi_2^\dagger \phi_2)^2
\nonumber
\\
&&
+ \lambda_3(\phi_1^\dagger \phi_1)
(\phi_2^\dagger \phi_2)
+ \lambda_4(\phi_1^\dagger \phi_2)
(\phi_2^\dagger \phi_1)
+ \frac{1}{2}
[
\lambda_5~(\phi_1^\dagger \phi_2)^2
+ ~{\rm h.c}
].
\end{eqnarray}
In the present work, the CP-conserving version of the THDM is examined. Subsequently, the parameters
in the above potential are set to be real. Additionally, the scalar potential of the THDM respects a $Z_2$ symmetry, i.e., $\phi_1 \leftrightarrow \phi_1$ and $\phi_2 \leftrightarrow -\phi_2$,
except for the soft-breaking terms
such as $m_{12}^2 \phi_1^\dagger \phi_2 + \text{h.c.}$,
which are added to the potential. Here, the parameter $m_{12}^2$ characterizes the soft breaking scale of the
$Z_2$ symmetry.

Two scalar doublets are
expanded around their
VEVs after electroweak symmetry breaking (EWSB) as
\begin{eqnarray}
\phi_k &=&
\begin{bmatrix}
\rho_k^+ \\
(v_k
+
\eta_k
+
i
\xi_k)
/\sqrt{2}
\end{bmatrix}
\quad \textrm{for} \quad
k=1,2.
\label{representa-htm}
\end{eqnarray}
The vacuum expectation value is then fixed at $v = \sqrt{v_1^2 + v_2^2} \sim 246$ GeV, in agreement with the SM value. After the EWSB, the physical particles in the THDM consist of two CP-even Higgs bosons, $h$ and $H$, with one of them, $h$, being identified as the SM-like Higgs boson discovered at the LHC, a CP-odd Higgs boson ($A^0$), and a pair of charged Higgs bosons ($H^\pm$). To obtain the physical masses of the new scalar bosons, we perform the following rotations:
\begin{eqnarray}
\begin{pmatrix}
\eta_1\\
\eta_2
\end{pmatrix}
&=&
\begin{pmatrix}
c_{\alpha}
&
-s_{\alpha}
\\
s_{\alpha}
&
c_{\alpha}
\end{pmatrix}
\begin{pmatrix}
H\\
h
\end{pmatrix},
\\
\begin{pmatrix}
\rho_1^{\pm}\\
 \rho_2^{\pm}
\end{pmatrix}
&=&
\begin{pmatrix}
c_{\beta}
&
-s_{\beta}
\\
s_{\beta}
&
c_{\beta}
\end{pmatrix}
\begin{pmatrix}
G^{\pm}\\
H^{\pm}
\end{pmatrix},
\\
\begin{pmatrix}
\xi_1\\
 \xi_2
\end{pmatrix}
&=&
\begin{pmatrix}
c_{\beta}
&
-s_{\beta}
\\
s_{\beta}
&
c_{\beta}
\end{pmatrix}
\begin{pmatrix}
G^{0}\\
A^0
\end{pmatrix}.
\end{eqnarray}
The mixing angle $\beta$ is defined by
$t_{\beta} \equiv \tan \beta = v_2/v_1$.
The physical masses of the Higgs bosons
are then expressed in terms of the model
parameters as follows:
\begin{eqnarray}
M_{H^{\pm}}^{2}
&=&
M^{2}-
\frac{1}{2}
\lambda_{45}
v^{2},
\\
M_{A^0}^{2}  &=&
M^{2}-\lambda_{5}v^{2},
\\
M_{h}^{2} &=& M_{11}^{2}
s_{\beta-\alpha}^{2}
+ M_{22}^{2}c_{\beta-\alpha}^{2}
+M_{12}^{2}s_{2(\beta-\alpha)},
\\
M_{H}^{2} &=&
M_{11}^{2}c_{\beta-\alpha}^{2}
+M_{22}^{2}s_{\beta-\alpha}^{2}
-M_{12}^{2}s_{2(\beta-\alpha)}
\end{eqnarray}
where the parameter $M^2$
is used as
$M^{2}=m_{12}^{2}/
(s_{\beta}c_{\beta})$.
The elements $M_{ij}$
for $i,j =1,2$ are given
by
\begin{eqnarray}
M_{11}^{2}&=&
(\lambda_{1}c_{\beta}^{4}
+\lambda_{2}s_{\beta}^{4})v^{2}
+\frac{v^{2}}{2}\;
\lambda_{345}
\; s_{2\beta}^{2}, \\
M_{22}^{2}
&=&
M^{2}
+ \frac{v^{2}}{4}
\Big[
\lambda_{12}
-2\lambda_{345}
\Big]
s_{2\beta}^{2}, \\
M_{12}^{2} &=&
M_{21}^{2}  =
-\frac{v^{2}}{2}
\Big[
\lambda_{1}c_{\beta}^{2}
-\lambda_{2}s_{\beta}^{2}
-
\lambda_{345}\; c_{2\beta}
\Big]
s_{2\beta}.
\end{eqnarray}
Here, the shorten notation have used
as $\lambda_{ij\cdots} = \lambda_{i}
+\lambda_{j}+\cdots$.

We present the relevant couplings involved in the amplitude computations for the processes $\gamma \gamma \rightarrow \phi_i\phi_j$ in Tables~\ref{THDM-coupling1},~\ref{THDM-coupling2} (for physical couplings) and in Table~\ref{THDM-coupling3} (for unphysical couplings). These couplings are derived in Appendix B.
\begin{table}[H]
\centering
{\begin{tabular}{
l@{\hspace{0.7cm}}
l@{\hspace{0.7cm}}
l}
\hline \hline
\textbf{Vertices}
&\textbf{Notations}
& \textbf{Couplings}\\
\hline \hline
$hW_{\mu}W_{\nu}$
& $g_{hWW}
\cdot
g_{\mu\nu}
$
&
$ i
\left(
\dfrac{2 M_W^2 }{v}
\; s_{\beta-\alpha}
\right)
\cdot
g_{\mu\nu}
$ \\
\hline
$HW_{\mu}W_{\nu}$
&
$g_{HWW}
\cdot
g_{\mu\nu}
$
&
$i
\left(
\dfrac{2 M_W^2 }{v}
\; c_{\beta-\alpha}
\right)
\cdot
g_{\mu\nu}
$
\\ \hline
$hZ_{\mu}Z_{\nu}$
&
$g_{hZZ}
\cdot
g_{\mu\nu}
$
&
$
i
\left(
\dfrac{2M_Z^2 }{v}
\; s_{\beta-\alpha}
\right)
\cdot
g_{\mu\nu}
$
\\
\hline
$HZ_{\mu}Z_{\nu}$
& $g_{HZZ}
\cdot
g_{\mu\nu}
$
&
$i
\left(
\dfrac{2M_Z^2 }{v}
\; c_{\beta-\alpha}
\right)
\cdot
g_{\mu\nu}
$ \\
\hline
$hH^{\pm}H^{\mp}$
& $
g_{hH^{\pm}H^{\mp}}
$
&
$
i
\Bigg[
\dfrac{
c_{\alpha+\beta}
(4M^2-3M_h^2 - 2M_{H^\pm}^2)
}
{2vs_{2\beta}}
$
\\
&&
\hspace{2cm}
$
+
\dfrac{
(2M_{H^\pm}^2-M_h^2)
c_{(\alpha-3\beta)}
}
{2vs_{2\beta}}
\Bigg]
$
\\
\hline
$Z_{\mu}H^{\pm}
(p_1)H^{\mp}(p_2)$
& $g_{ZH^{\pm}H^{\mp}}
\cdot
(p_1- p_2)_{\mu}
$
&
$
i
\left(
\dfrac{M_Z\; c_{2W}}{v}
\right)
\cdot
(p_1 - p_2)_{\mu}$
\\
\hline
$A_{\mu} H^{\pm}(p_1)
H^{\mp}(p_2)$
& $g_{A H^{\pm}H^{\mp}}
\cdot
(p_1- p_2)_{\mu}
$
&
$
i
\left(
\dfrac{
M_Z\; s_{2W}
}{v}
\right)
\cdot
(p_1- p_2)_{\mu}
$
\\
\hline
$hhh$
& $g_{hhh} $
&
$
i
\dfrac{3e}
{4M_W s_W
s_{2\beta}}
\Bigg[
M^2c_{\alpha-3\beta}
+(M^2 - M_h^2)
c_{3\alpha-\beta}
$
\\
&&
\hspace*{4cm}
$
+(2M^2-3 M_h^2)
c_{\alpha+\beta}
\Bigg]
$
\\
\hline
$HHH$
& $g_{HHH} $
&
$
i
\dfrac{3e}{4M_Ws_Ws_{2\beta}}
\Bigg[
M^2s_{\alpha-3\beta}
+(M_H^2-M^2)s_{3\alpha-\beta}
$
\\
&&
\hspace{4cm}
$
+(2M^2-3M_H^2)s_{\alpha+\beta}
\Bigg]
$
\\
\hline
$hHH$
& $g_{hHH}$
&
$
i
\dfrac{
\Big[
s_{2\alpha}(3M^2-M_h^2-2M_H^2)
+
M^2s_{2\beta}
\Big]
}{v\; s_{2\beta}}
s_{\alpha-\beta}
$
\\
\hline
$Hhh$
&
$g_{Hhh}$
&
$
i
\dfrac{
\Big[
s_{2\alpha}(3M^2-M_H^2-2m_h^2)
-M^2s_{2\beta}
\Big]
}{2vs_{2\beta}}
c_{\alpha-\beta}
$
\\
\hline
\hline
\end{tabular}}
\caption{
\label{THDM-coupling1}
The physical couplings 
contributing to the
considered processes
$\gamma \gamma\rightarrow
\phi_i\phi_j$
in the THDM. }
\end{table}
\begin{table}[H]
\centering
{\begin{tabular}{
l@{\hspace{1.2cm}}
l@{\hspace{1.2cm}}
l}
\hline \hline
\textbf{Vertices}
&\textbf{Notations}
& \textbf{Couplings}
\\
\hline
\hline
$H H^{\pm}H^{\mp}$
& 
$g_{H H^{\pm}H^{\mp}}$
&
$
i
\Bigg[
\dfrac{
s_{\alpha+\beta}
(4M^2-3M_H^2-2M_{H^\pm}^2)
}
{2v\; s_{2\beta}}
$
\\
&
&
\hspace{2cm}
$
+
\dfrac{
(2M_{H^\pm}^2 - M_H^2)
s_{\alpha-3\beta}}
{2v\; s_{2\beta}}
\Bigg]
$
\\
\hline
$H(p_1) H^{\pm}
(p_2)
W^{\mp}_{\mu}$
& $g_{H H^{\pm} W}
\cdot
(p_1-p_2)_{\mu}
$
&
$
\pm i
\left(
\dfrac{M_W
}{v} 
s_{\beta-\alpha}
\right)
\cdot
(p_1-p_2)_{\mu}
$
\\
\hline
$h(p_1) H^{\pm}(p_2)
W^{\mp}_{\mu}$
&
$g_{h H^{\pm} W}
\cdot
(p_1-p_2)_{\mu}
$
&
$
\mp i
\left(
\dfrac{M_W}{v}
c_{\alpha-\beta}
\right)
\cdot
(p_1-p_2)_{\mu}
$
\\
\hline
$H^{\pm} H^{\mp} A_{\mu}A_{\nu}$
&
$g_{H^{\pm} H^{\mp}AA}
\cdot
g_{\mu\nu}
$
&
$
i
\left(
\dfrac{4 M_W^2 s_W^2}{v^2}
\right)
\cdot
g_{\mu\nu}
$
\\
\hline
$H H^{\pm} W^{\mp}_{\mu}
A_{\nu}$
&
$g_{H H^{\pm} W A}
\cdot
g_{\mu\nu}$
&
$
i
\left(
\dfrac{2 M_W^2 s_W
}{v^2}
s_{\alpha-\beta}
\right)
\cdot
g_{\mu\nu}
$
\\
\hline
$hH^{\pm}
W^{\mp}_{\mu}A_{\nu}$
&
$
g_{h H^{\pm} W A}
\cdot
g_{\mu\nu}
$
&
$ i
\left(
\dfrac{
2 M_W^2 s_W
}{v^2}
c_{\alpha-\beta}
\right)
\cdot
g_{\mu\nu}
$
\\
\hline
$hH H^{\pm}H^{\mp}$
&
$g_{HhH^{\pm}H^{\mp}}$
&
$i
\lambda_{HhH^{\pm}H^{\mp}}
$
[in Eq.~(\ref{gHhSS})]
\\
\hline
$HH H^{\pm}H^{\mp}$
& $g_{HH H^{\pm}H^{\mp} } $
&
$i \lambda_{HH H^{\pm}H^{\mp}}$
[in Eq.~(\ref{gHHHpHm})]
\\
\hline
$hhH^{\pm}H^{\mp}$
&
$g_{hhH^{\pm}H^{\mp} } $
&
$i \lambda_{hh
H^{\pm}H^{\mp}}$
[in Eq.~(\ref{ghhHpHm})]
\\
\hline
$hh W^{\pm}_{\mu}W^{\mp}_{\nu}$
& $g_{hhWW}\cdot g_{\mu\nu} $
&
$
i
\left(
\dfrac{4M_W^2}{v^2}
\right)
\cdot g_{\mu\nu}
$
\\
\hline
$HH W^{\pm}_{\mu}W^{\mp}_{\nu}$
&
$g_{HH WW }\cdot g_{\mu\nu}$
&
$i
\left(
\dfrac{4M_W^2}{v^2}
\right)
\cdot g_{\mu\nu}
$
\\
\hline
\hline
\end{tabular}}
\caption{
\label{THDM-coupling2}
Additional, 
the physical couplings 
contributing to the 
considered processes
$\gamma \gamma\rightarrow
\phi_i\phi_j$
in the THDM.}
\end{table}
\begin{table}[H]
\centering
{\begin{tabular}{
l@{\hspace{1.2cm}}
l@{\hspace{1.2cm}}
l}
\hline \hline
\textbf{Vertices}
&\textbf{Notations}
& \textbf{Couplings}
\\
\hline
\hline
$A_{\mu}W^{\pm}_{\nu}G^{\mp}$
&
$
g_{AW^{\pm}G^{\mp}}
\cdot g_{\mu\nu}$
&
$
i
\left(
\dfrac{2M_W^2s_W}{v}
\right)
\cdot g_{\mu\nu}
$
\\
\hline
$HH^{\pm}G^{\mp}$
&
$g_{HH^{\pm}G^{\mp}}$
&
$i
\left(
\dfrac{e}
{2M_Ws_W}
 s_{\alpha-\beta}
\right)
(M_{H^\pm}^2-M_H^2)
$
\\
\hline
$A_{\mu}A_{\nu}
G^{\pm}G^{\mp}$
&
$g_{AAG^{\pm}G^{\mp}}
\cdot g_{\mu\nu}$
&
$i
\left(
\dfrac{4M_W^2s_W^2}{v^2}
\right)
\cdot g_{\mu\nu}
$
\\
\hline
$A_{\mu}
G^{\pm}(p_1)
G^{\mp}(p_2)
$
&
$g_{AG^{\pm}G^{\mp}}
\cdot
(p_1-p_2)_{\mu}
$
&
$ i
\left(
\dfrac{2M_Ws_W}{v}
\right)
\cdot
(p_1-p_2)_{\mu}$
\\
\hline\hline
$hHG^{\pm}G^{\mp}$
&
$g_{hHG^{\pm}G^{\mp} } $
&
$i\lambda_{hHG^{\pm}G^{\mp}} $
[in Eq.~(\ref{hHGG})]
\\
\hline
$hhG^{\pm}G^{\mp}$
&
$g_{hh G^{\pm}G^{\mp} } $
&
$i\lambda_{hhG^{\pm}G^{\mp}} $
[in Eq.~(\ref{hhGpGm}) ]
\\
\hline
$HHG^{\pm}G^{\mp}$
&
$g_{HHG^{\pm}G^{\mp} } $
&
$i\lambda_{HHG^{\pm}G^{\mp}} $
[in Eq.~(\ref{HHGpGm})]
\\
\hline
\hline
\end{tabular}}
\caption{
\label{THDM-coupling3}
Unphysical couplings 
involving to the processes
under investigations
are shown.}
\end{table}
Finally, we pay attention to the 
Yukawa sector in the THDM. 
In order to avoid Tree-level
Flavor-Changing Neutral
Currents (FCNCs), the 
discrete $Z_2$-symmetry
may be proposed in the THDM as 
in~\cite{Aoki:2009ha}. 
The $Z_2$-parity assignments for 
all fermions and the definition for
four types of the THDM 
based on the couple of
the scalar with fermions
are shown in~\cite{Phan:2024zus}. 
The Yukawa Lagrangian is then
written in the mass
eigenstates as
in~\cite{Branco:2011iw}
\begin{eqnarray}
\label{YukawaTHDM}
{\mathcal{L}}_{\rm Yukawa} =
-\sum_{f=u,d,\ell}
\left(g_{hff}\cdot
\bar{f} f h +
g_{Hff}\cdot  \bar{f} f H
- i g_{A^0ff}\cdot
\bar{f} \gamma_5 f A^0 \right)
+ \cdots,
\end{eqnarray}

We show the Yukawa couplings of CP-even with 
fermions for four types of the THDM
in Table~\ref{YukawaTHDM},
see~\cite{Branco:2011iw, Phan:2024jbx}
for further detail.
\begin{table}[H]
\centering
\begin{tabular}
{
l
lll
lll
}
\hline\hline
Type
&$g_{huu}$
&$g_{hdd}$
&$g_{h\ell\ell}$
&$g_{Huu}$
&$g_{Hdd}$
&$g_{H\ell\ell}$
\\
\hline
\hline
&&&
\\
I
&
$\dfrac{m_u}{\sqrt{2}v}
\dfrac{c_\alpha}{s_\beta}$
&
$\dfrac{m_d}{\sqrt{2}v}
\dfrac{c_\alpha}{s_\beta}$
&
$\dfrac{m_{\ell}}{\sqrt{2}v}
\dfrac{c_\alpha}{s_\beta}$
&
$\dfrac{m_u}{\sqrt{2}v}
\dfrac{s_\alpha}{s_\beta}$
&
$\dfrac{m_d}{\sqrt{2}v}
\dfrac{s_\alpha}{s_\beta}$
&
$\dfrac{m_{\ell}}{\sqrt{2}v}
\dfrac{s_\alpha}{s_\beta}$
\\
\hline
&&&
\\
II
&
$\dfrac{m_u}{\sqrt{2}v}
\dfrac{c_\alpha}{s_\beta}$
&
$-\dfrac{m_d}{\sqrt{2}v}
\dfrac{s_\alpha}{c_\beta}$
&
$-\dfrac{m_{\ell}}{\sqrt{2}v}
\dfrac{s_\alpha}{c_\beta}
$
&
$\dfrac{m_u}{\sqrt{2}v}
\dfrac{s_\alpha}{s_\beta}$
&
$-\dfrac{m_d}{\sqrt{2}v}
\dfrac{c_\alpha}{c_\beta}$
&
$-\dfrac{m_{\ell}}{\sqrt{2}v}
\dfrac{c_\alpha}{c_\beta}
$
\\
\hline
&&&
\\
X
&
$\dfrac{m_u}{\sqrt{2}v}
\dfrac{c_\alpha}{s_\beta}$
&
$\dfrac{m_d}{\sqrt{2}v}
\dfrac{c_\alpha}{s_\beta}$
&
$-\dfrac{m_{\ell}}{\sqrt{2}v}
\dfrac{s_\alpha}{c_\beta}$
&
$\dfrac{m_u}{\sqrt{2}v}
\dfrac{s_\alpha}{s_\beta}$
&
$\dfrac{m_d}{\sqrt{2}v}
\dfrac{s_\alpha}{s_\beta}$
&
$-\dfrac{m_{\ell}}{\sqrt{2}v}
\dfrac{c_\alpha}{c_\beta}$
\\
\hline
&&&
\\
Y
&
$\dfrac{m_u}{\sqrt{2}v}
\dfrac{c_\alpha}{s_\beta}$
&
$-\dfrac{m_d}{\sqrt{2}v}
\dfrac{s_\alpha}{c_\beta}$
&
$\dfrac{m_{\ell}}{\sqrt{2}v}
\dfrac{c_\alpha}{s_\beta}$
&
$\dfrac{m_u}{\sqrt{2}v}
\dfrac{s_\alpha}{s_\beta}$
&
$-\dfrac{m_d}{\sqrt{2}v}
\dfrac{c_\alpha}{c_\beta}$
&
$\dfrac{m_{\ell}}{\sqrt{2}v}
\dfrac{s_\alpha}{s_\beta}$
\\
\hline
\hline
\end{tabular}
\caption{\label{YukawaTHDM}
We show all the Yukawa
couplings of CP-even Higges 
to fermions for all types of THDMs.
}
\end{table}
The parameter space
$\mathcal{P}_{\rm THDM}$ for THDM
is include as follows
\begin{eqnarray}
\label{thdmspace}
	\mathcal{P}_{\rm THDM} =
	\{M_h^2\sim 125. \textrm{GeV}
	, M_H^2, M_{A^0}^2,
	M^2_{H^{\pm}}, m_{12}^2,
	t_{\beta}, s_{\beta-\alpha} \}.
\end{eqnarray}
As in the case of the IHDM, we first summarize the current constraints on the parameter space of the THDM, given in Eq.~\ref{thdmspace}. Both theoretical conditions and experimental data are taken into account, from which the allowed regions of the parameter space of the THDM are determined. Theoretical constraints arise from the requirements that the model remains within the perturbative regime, satisfies tree-level unitarity, and ensures vacuum stability of the scalar potential. These topics have been discussed in detail in Refs.~\cite{Nie:1998yn, Kanemura:1999xf, Akeroyd:2000wc, Ginzburg:2005dt, Kanemura:2015ska} and references therein.
We also consider electroweak precision tests (EWPT) in the context of the THDM, based on experimental data. Relevant implications for these constraints from LEP are reported in Refs.~\cite{Bian:2016awe, Xie:2018yiv}. The allowed mass ranges of the scalar particles in the THDM have been investigated at LEP, the Tevatron, and the LHC, as reviewed in Ref.~\cite{Kanemura:2011sj}. Furthermore, one-loop induced decays of the SM-like Higgs boson, such as $h \rightarrow \gamma\gamma$ and $h \rightarrow Z\gamma$, have also been analyzed within the THDM in Refs.~\cite{Chiang:2012qz, Benbrik:2022bol} and references therein.
By combining all of the above constraints, the physical parameter ranges can be chosen as follows: $126~\text{GeV} \leq M_H \leq 1000~\text{GeV}$, $60~\text{GeV} \leq M_{A^0} \leq 1000~\text{GeV}$, and $80~\text{GeV} \leq M_{H^{\pm}} \leq 1000~\text{GeV}$ for Type-I and Type-X THDMs. For Type-II and Type-Y, the physical parameters are scanned in the ranges: $500~\text{GeV} \leq M_H \leq 1000~\text{GeV}$, $500~\text{GeV} \leq M_{A^0} \leq 1000~\text{GeV}$, and $580~\text{GeV} \leq M_{H^{\pm}} \leq 1500~\text{GeV}$. The soft $Z_2$-breaking parameter is chosen as $m_{12}^2 = M_H^2 s_\beta c_\beta$.
Finally, additional constraints in the $(t_\beta, M_{H^\pm})$ plane are examined by incorporating flavor physics data, as shown in Ref.~\cite{Haller:2018nnx}. The results in Ref.~\cite{Haller:2018nnx} indicate that small values of $t_\beta$ are favored to be consistent with flavor experimental constraints. In our complementary analysis, we also explore the phenomenology associated with low $t_\beta$ scenarios.
\section{Phenomenological results}
The detailed calculations for
$\gamma \gamma \rightarrow \phi_i
\phi_j$ with CP-even Higgses
$\phi_{i,j} \equiv h, H$ in the
HESM have been reported in our previous
work~\cite{Phan:2024vxm}. In this
paper, we focus on the phenomenological
analysis of
the concerned processes
in the IHDM and the THDM.
We are interested in examining
the enhancement factors
$\mu_{\phi_i\phi_j}^{\textrm{NP}}$
with NP standing for the THDM and the IHDM,
respectively, defined as
the ratio of the cross-sections
of $\gamma \gamma \rightarrow\phi_i\phi_j$
in the HESM to the corresponding ones
for $\gamma \gamma
\rightarrow hh$ in the SM. The factors
are given explicitly by
\begin{eqnarray}
\label{EF_factors}
\mu_{\phi_i\phi_j}^{\textrm{NP}}=
\frac{\hat{\sigma}^{\textrm{NP}}_{\gamma \gamma
\rightarrow \phi_i\phi_j}
}{
\hat{\sigma}^{\textrm{SM}}_{\gamma \gamma
\rightarrow hh}
}
(\mathcal{P}_{\textrm{NP}}).
\end{eqnarray}
In this work, the enhancement
factors are examined within
the parameter space
of the THDM and the IHDM.
Forthe phenomenological results, all
physical input parameters in the SM
are taken to be the same as those
in~\cite{Phan:2024zus,Phan:2024jbx}.
\subsection{IHDM}
Phenomenological studies for
the processes $\gamma \gamma
\rightarrow \phi_i \phi_j$
in the IHDM are presented in this
subsection. In the IHDM, the process
$\gamma \gamma \rightarrow hH$
is forbidden by the $Z_2$-symmetry.
For this reason, we only
focus on
physical results for the
processes $\gamma \gamma
\rightarrow hh,~HH$ in the IHDM.
We note that all physical results
in the IHDM presented in the
following subsections can be
considered
as the first outcomes
from this work.
\subsubsection{Production
cross-sections}
\begin{figure}[H]
\centering
\begin{tabular}{cc}
\\
\hspace{-3cm}
$\hat{\sigma}_{\phi_i \phi_j}(\mu_2^2=0)$ [pb]
&
\hspace{-2.5cm}
$\hat{\sigma}_{\phi_i \phi_j} (\mu_2^2=200^2)$
[pb]
\\
\includegraphics[width=8cm, height=6cm]{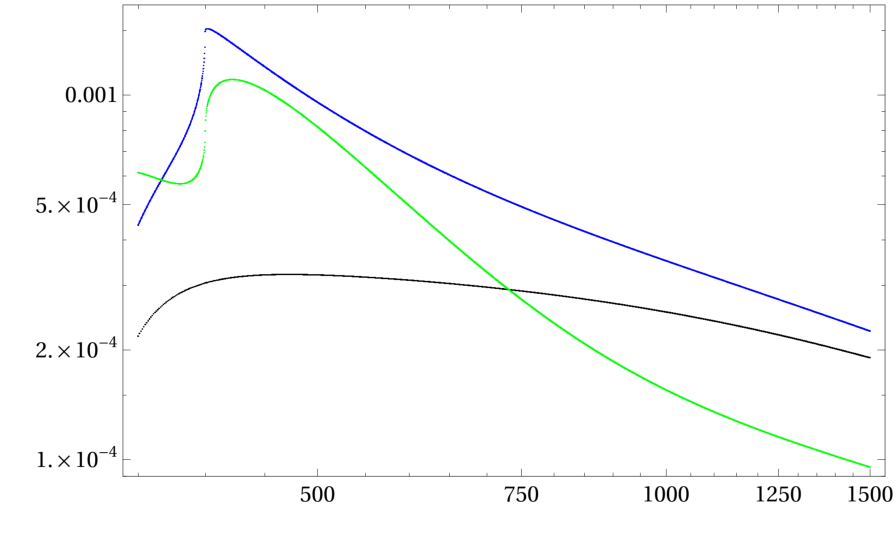}
&
\includegraphics[width=8cm, height=6cm]{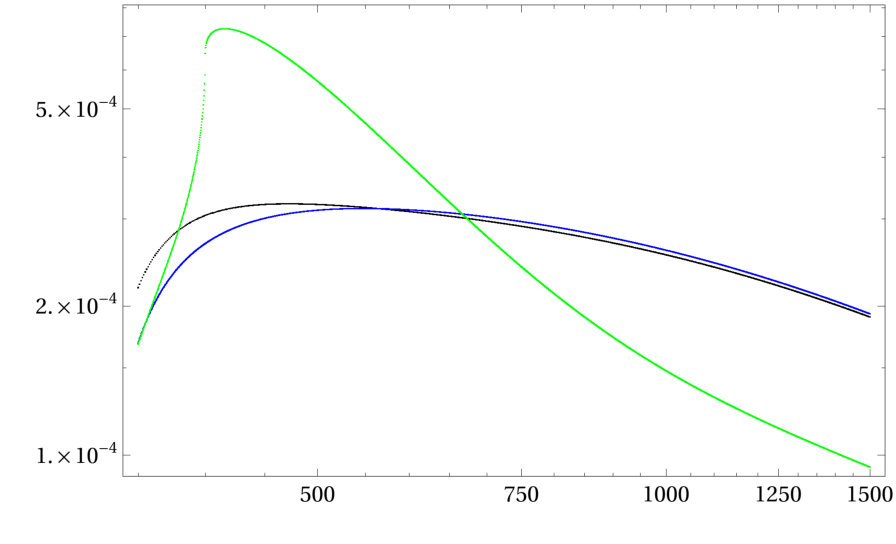}
\\
\hspace{6cm}
$\sqrt{\hat{s}_{\gamma\gamma}}$ [GeV]
&
\hspace{6cm}
$\sqrt{\hat{s}_{\gamma\gamma}}$ [GeV]
\end{tabular}
\caption{\label{totalIHDM}
Total cross-sections for
$\gamma \gamma \rightarrow hh,~HH$
in the SM and IHDM are presented
as functions of $\sqrt{\hat{s}_{\gamma\gamma}}$.
For the generated data, we select
$M_{H^{\pm}}=200$ GeV, $M_H=150$ GeV
in this case. We vary $350$ GeV $\leq
\sqrt{\hat{s}_{\gamma\gamma}}
\leq 1500$ GeV. In the plots below,
we fix $\mu_2^2 = 0, 200^2$ GeV$^2$
and $\lambda_2 = 0.8$ for all cases.
In the plots,
the black line shows
$\hat{\sigma}_{hh}$ in the SM.
Additionally, the blue (green)
line presents $\gamma \gamma
\rightarrow hh~(HH)$ in the THDM.
}
\end{figure}

In Fig.~\ref{totalIHDM},
we show cross-sections for
$\gamma \gamma \rightarrow hh,~HH$
in the IHDM, together with those
for $hh$ production in the SM,
as functions of center-of-mass energy
(CoM, or $\sqrt{\hat{s}_{\gamma\gamma}}$).
For the generated data,
we select the following
parameter space in the IHDM:
$M_{H^{\pm}} = 200$ GeV,
$M_H = 150$ GeV, and fix $\lambda_2 = 0.8$
for all cases. We vary $350$ GeV $\leq
\sqrt{\hat{s}_{\gamma\gamma}} \leq 1500$ GeV
in the plots. Cross-sections are presented
for $\mu_2^2 = 0$ GeV$^2$ on the left panel
and for $200^2$ GeV$^2$ on the right
panel, respectively. In the plots, the black line
represents the cross-sections for
$\gamma \gamma \rightarrow hh$ in the SM,
and the blue (green) line
represents the cross-sections
for $\gamma \gamma \rightarrow hh,~(HH)$
in the IHDM, respectively.

We first comment on the results
in the case of $\mu_2^2=0$ GeV$^2$.
The cross-sections for $hh,~HH$ in
the IHDM have peaks at
$\sqrt{\hat{s}_{\gamma\gamma}}
\sim 2 M_{H^{\pm}}=400$ GeV.
In the regions
$\sqrt{\hat{s}_{\gamma\gamma}}
\leq 750$ GeV,
$\hat{\sigma}_{hh,HH}$ in the IHDM
are larger than
$\hat{\sigma}_{hh}$ in the SM.
Beyond the regions of
$\sqrt{\hat{s}_{\gamma\gamma}} \geq 750$ GeV,
the cross-sections for $HH$ in the IHDM
are suppressed in comparison with $hh$
production in both the SM and the IHDM.
It is interesting to find that
the production cross-sections
for $\gamma \gamma
\rightarrow hh,~HH$ in the IHDM
are dominant around the peaks
compared with $\hat{\sigma}_{hh}$ in the SM.
This indicates
that the contributions from
singly charged Higgs in the loop of
$\gamma \gamma \rightarrow hh,~HH$
are significant in these regions.

In the case of $\mu_2^2=200^2$ GeV$^2$,
we only observe a peak of
$\hat{\sigma}_{HH}$ in the IHDM around
$\sqrt{\hat{s}_{\gamma\gamma}}
\sim 2M_{H^{\pm}} =400$ GeV.
Furthermore,
the data shows that the cross-sections
for $HH$ production are dominant in the
regions $\sqrt{\hat{s}_{\gamma\gamma}}
\leq 750$ GeV, in contrast with
the corresponding ones for $hh$
production in both the SM and the IHDM.
Beyond the regions
$\sqrt{\hat{s}_{\gamma\gamma}}
\geq 750$ GeV, $\hat{\sigma}_{HH}$
is suppressed.
It is observed
that the cross-sections for
$hh$ production
in the IHDM are smaller than
$\hat{\sigma}_{hh}$ in the SM
when $\sqrt{\hat{s}_{\gamma\gamma}}
\leq 550$ GeV. In the regions of
$\sqrt{\hat{s}_{\gamma\gamma}} \geq 550$
GeV, $\hat{\sigma}_{hh}$ in the IHDM
tends to the
cross-sections for $hh$ production
in the SM. This can be explained as follows.
Since the $hh$ production
in the IHDM are different from those
in the SM by the contributions of charged Higgs
in the loop of triangle $h^*$-pole
and box diagrams. These contributions
depend on $M_{H^{\pm}}$
and the vertices $h H^{\pm}H^{\mp}$,
$hh H^{\pm}H^{\mp}$, expressed in terms of
$\mu_2^2$. At large values of
$\mu_2^2$, these contributions
may be cancelled out. As a result,
cross-sections for $hh$ production
in the IHDM tend to the
corresponding ones in the SM.
In another case of $HH$ production,
we have no couplings
of $H H^{\pm}H^{\mp}$ due to the $Z_2$-symmetry
and the vertex $HH H^{\pm}H^{\mp}$
depends on $\lambda_2$. Therefore, we have no
such cancellations as mentioned.
It is reasonable
that the cross-sections
for $HH$ production
in the IHDM are dominant in the regions
$\sqrt{\hat{s}_{\gamma\gamma}}
\leq 550$ GeV and they also have a peak at
$\sqrt{\hat{s}_{\gamma\gamma}}
\sim 2 M_{H^{\pm}} = 400$ GeV.
\subsubsection{Enhancement factors} 
\begin{figure}[H]
\centering
\begin{tabular}{cc}
\hspace{-6.2cm}
$\mu_{hh}^{\textrm{IHDM}}$
&
\hspace{-6.2cm}
$\mu_{hh}^{\textrm{IHDM}}$
\\
\includegraphics[width=8cm, height=6cm]
{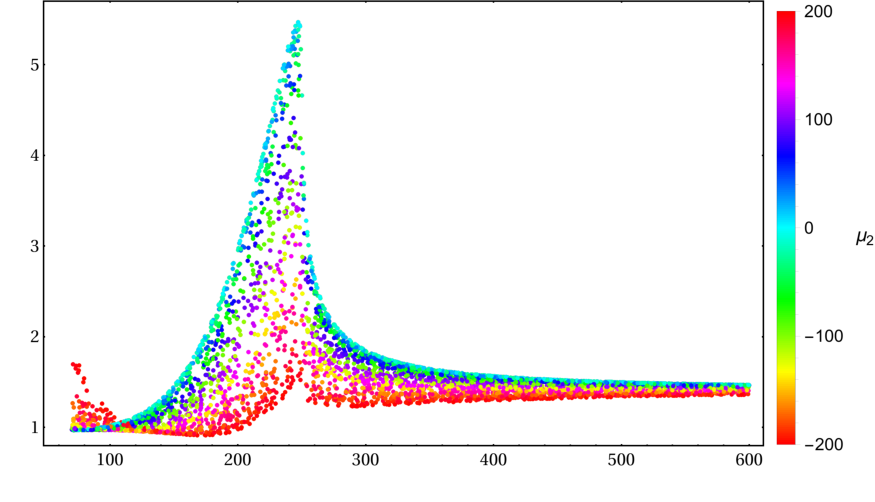}
&
\includegraphics[width=8cm, height=6cm]
{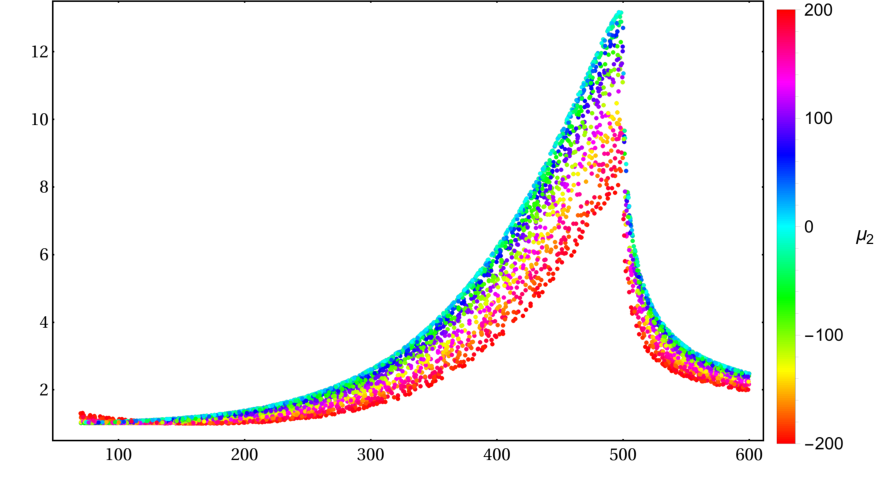}
\\
\hspace{5cm}
$M_{H^{\pm}}$ [GeV]
&
\hspace{5cm}
$M_{H^{\pm}}$
[GeV]
\\
&
\\
\hspace{-6cm}
$\mu_{HH}^{\textrm{IHDM}}$
&
\hspace{-6cm}
$\mu_{HH}^{\textrm{IHDM}}$
\\
\includegraphics[width=8cm, height=6cm]
{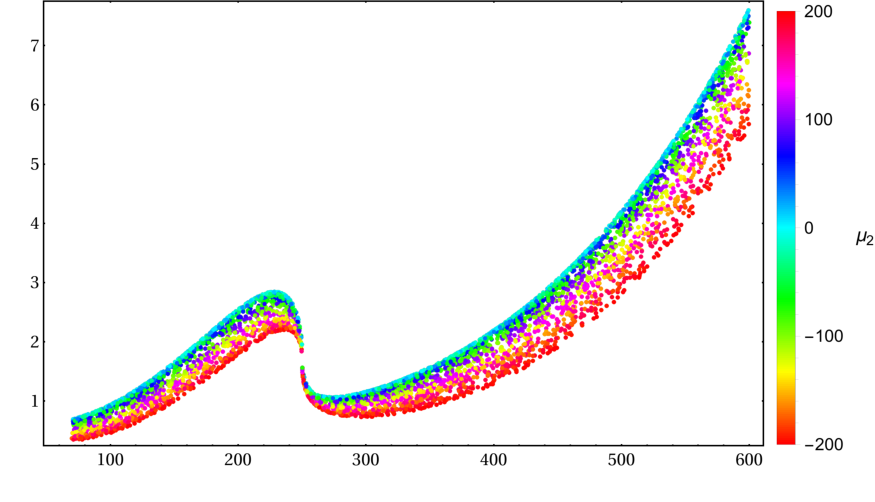}
&
\includegraphics[width=8cm, height=6cm]
{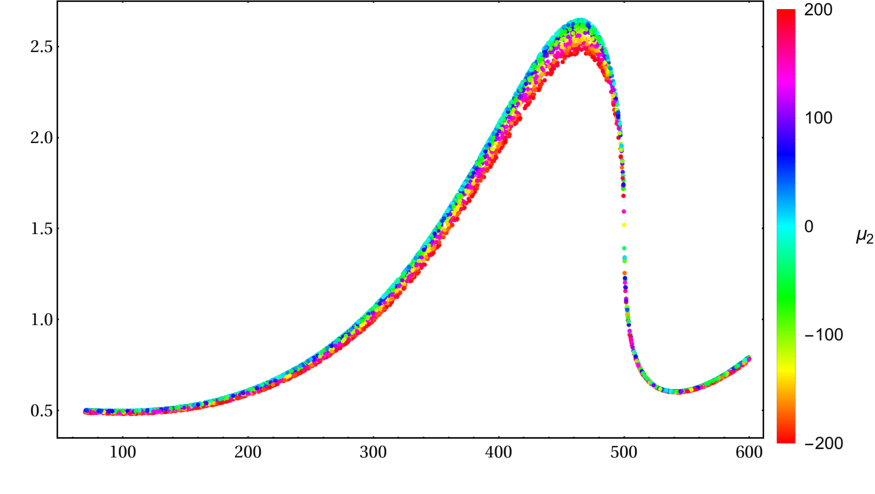}
\\
\hspace{5cm}
$M_{H^{\pm}}$ [GeV]
&
\hspace{5cm}
$M_{H^{\pm}}$
[GeV]
\end{tabular}
\caption{\label{fig_mu_IHDM}
The enhancement factors are presented in the parameter space of $M_{H^{\pm}},~\mu_2^2$. Charged Higgs masses are varied as $70$ GeV $\leq M_{H^{\pm}} \leq 1000$ GeV and $-200$ GeV $\leq \mu_2 \leq 200$ GeV. We fix $\lambda_2 = 0.8$ and $M_H = 150$ GeV for all cases. In the plots, we set $\sqrt{\hat{s}_{\gamma\gamma}} = 500$ GeV (for the left panel-plots) and $\sqrt{\hat{s}_{\gamma\gamma}} = 1000$ GeV (for the right panel-plots).
}
\end{figure}

The enhancement factors given
in Eq.~\ref{EF_factors}
are examined in the IHDM.
In Fig.~\ref{fig_mu_IHDM},
the factors for $\gamma \gamma
\rightarrow hh,~HH$ are scanned in
the parameter space of
$M_{H^{\pm}},~\mu_2^2$.
In the following scatter plots,
singly charged Higgs masses
are varied from
$70$ GeV $\leq M_{H^{\pm}}
\leq 600$ GeV and $-200$ GeV
$\leq \mu_2 \leq 200$ GeV.
Furthermore, we fix
$\lambda_2 = 0.8$ and
$M_H=150$ GeV for all cases.
The data are generated at
$\sqrt{\hat{s}_{\gamma\gamma}}=500$
(all left panel scatter-plots)
GeV and at $\sqrt{\hat{s}_{\gamma\gamma}}
=1000$ GeV (all right panel scatter-plots).

The factors for $hh$ production in the IHDM are first analyzed. Since the cross-sections for $hh$ production are enhanced around the peaks at $\sqrt{\hat{s}_{\gamma\gamma}} \sim 2 M_{H^{\pm}}$, it is not surprising to find that $\mu_{hh}^{\textrm{IHDM}}$ becomes largest at $\sqrt{\hat{s}_{\gamma\gamma}} \sim 2 M_{H^{\pm}} = 250$ GeV (for the left plots) and at $\sqrt{\hat{s}_{\gamma\gamma}} \sim 2 M_{H^{\pm}} = 500$ GeV (for the right plots). Around these peaks, the data indicate that the enhancement factors tend to about $\sim 1.5$ in the limit of $\mu_2^2 \rightarrow M_{H^{\pm}}^2$, since the contributions of the charged Higgs in the loop are small when $\mu_2^2 \rightarrow M_{H^{\pm}}^2$ (due to the fact that the couplings of $h H^{\pm} H^{\mp}$ and $hh H^{\pm} H^{\mp}$ tend to zero in this limit). It is found that the enhancement factors can reach a factor of $6$ (for $500$ GeV of CoM) and a factor of $13$ (for $1000$ GeV of CoM) around the peaks. Beyond the peaks, we observe that $1 \leq \mu_{hh}^{\textrm{IHDM}} \leq 2$.

For the enhancement factors of $HH$ production in the IHDM, we also find that $\mu_{HH}^{\textrm{IHDM}}$ becomes largest at
$\sqrt{\hat{s}_{\gamma\gamma}} \sim 2 M_{H^{\pm}} = 250$ GeV (for the left plots) and at $\sqrt{\hat{s}_{\gamma\gamma}} \sim 2 M_{H^{\pm}} = 500$ GeV (for the right plots). It is important to realize that $\mu_{HH}^{\textrm{IHDM}}$ has different behavior in comparison with $\mu_{hh}^{\textrm{IHDM}}$. At $500$ GeV of CoM, the factors rise to the peak and then decrease rapidly beyond the peak. However, they grow with the charged Higgs masses in the above regions of $M_{H^{\pm}} \geq \sim 300$ GeV. Because there are no couplings of $H H^{\pm} H^{\mp}$ due to the $Z_2$-symmetry and the vertex $HH H^{\pm} H^{\mp}$ only depends on $\lambda_2$, the factors $\mu_{HH}^{\textrm{IHDM}}$ increase with $M_{H^{\pm}}$ in the high regions of charged Higgs masses.

\begin{figure}[H]
\centering
\begin{tabular}{cc}
\hspace{-7cm}
$\mu_{hh}^{\textrm{IHDM}}$
&
\hspace{-7cm}
$\mu_{hh}^{\textrm{IHDM}}$
\\
\includegraphics[width=8cm, height=6cm]
{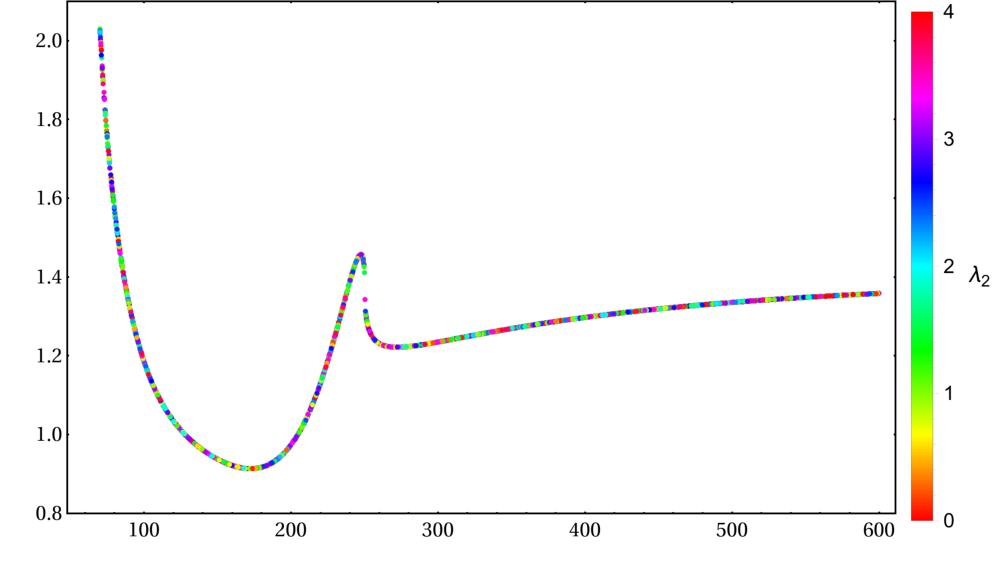}
&
\includegraphics[width=8cm, height=6cm]
{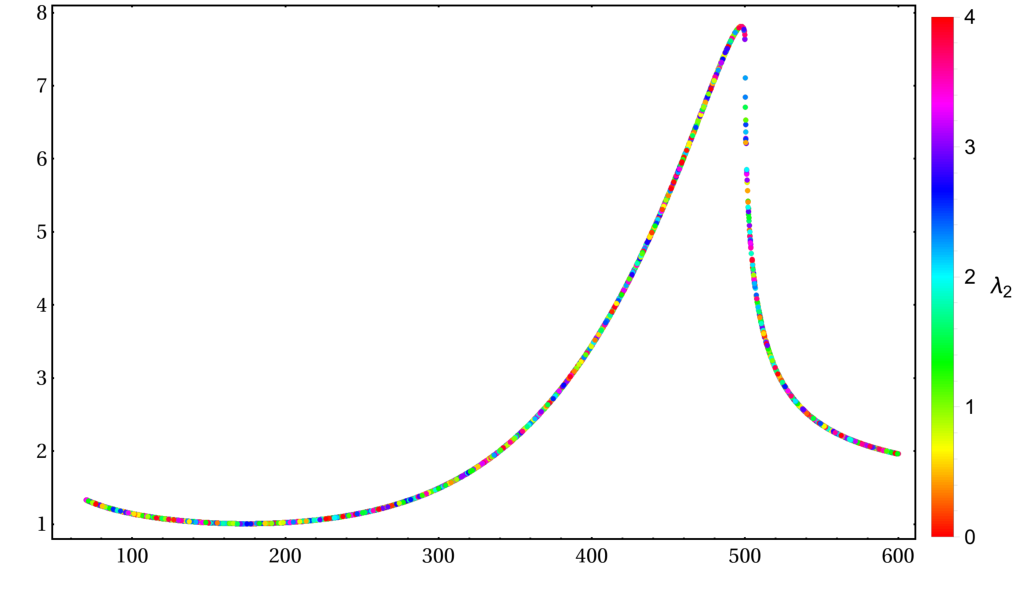}
\\
\hspace{5cm}
$M_{H^{\pm}}$ [GeV]
&
\hspace{5cm}
$M_{H^{\pm}}$
[GeV]
\\
&
\\
\hspace{-7cm}
$\mu_{HH}^{\textrm{IHDM}}$
&
\hspace{-7cm}
$\mu_{HH}^{\textrm{IHDM}}$
\\
\includegraphics[width=8cm, height=6cm]
{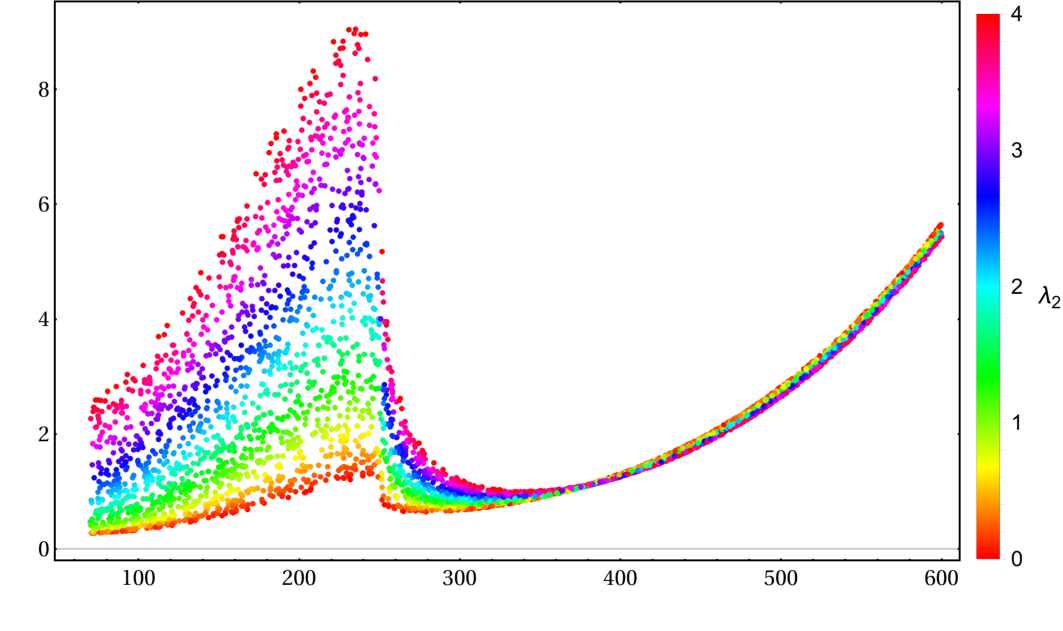}
&
\includegraphics[width=8cm, height=6cm]
{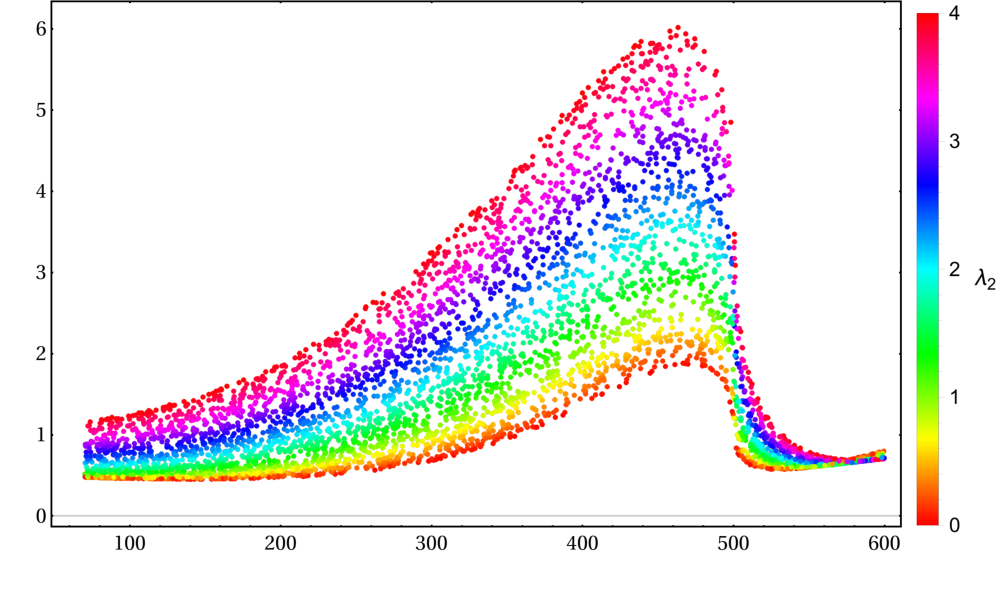}
\\
\hspace{5cm}
$M_{H^{\pm}}$ [GeV]
&
\hspace{5cm}
$M_{H^{\pm}}$
[GeV]
\end{tabular}
\caption{\label{fig_lam2_IHDM}
The enhancement factors are scanned over the parameter space of $M_{H^{\pm}}, \lambda_2$. Charged Higgs masses are in $70$ GeV $\leq M_{H^{\pm}} \leq 1000$ GeV and $0 \leq \lambda_2 \leq 4$. We fix $\mu_2^2 = 200^2$ GeV$^2$ and $M_H = 150$ GeV for all cases. In the plots, we set $\sqrt{\hat{s}_{\gamma\gamma}} = 500$ GeV (for the above figures) and $\sqrt{\hat{s}_{\gamma\gamma}} = 1000$ GeV (for the below figures), correspondingly.
}
\end{figure}

In Fig.~\ref{fig_lam2_IHDM}, the enhancement factors for $\gamma \gamma \rightarrow hh,~HH$ are generated in the space of $M_{H^{\pm}},~\lambda_2$. The charged Higgs masses are varied as $70$ GeV $\leq M_{H^{\pm}} \leq 600$ GeV and $0 \leq \lambda_2 \leq 4$. We fix $\mu_2^2 = 200^2$ GeV$^2$ and $M_H = 150$ GeV for all cases. In the scatter plots, we set $\sqrt{\hat{s}_{\gamma\gamma}} = 500$ (for all left panel plots) GeV and $\sqrt{\hat{s}_{\gamma\gamma}} = 1000$ GeV (for all right panel plots). For the factors $\mu_{hh}^{\textrm{IHDM}}$ (as shown in all the above scatter plots), both the couplings $h H^{\pm}H^{\mp}$ and $hh H^{\pm}H^{\mp}$ are independent of $\lambda_2$. As a result, the factors only depend on $M_{H^{\pm}}$. For the factors $\mu_{HH}^{\textrm{IHDM}}$ (as shown in all the below scatter plots), it is found that the quadratic-coupling $HH H^{\pm}H^{\mp}$ depends on $\lambda_2$. As a result, the factors depend strongly on $\lambda_2$ and $M_{H^{\pm}}$. These massive contributions are mainly from the charged Higgs exchanging in the box diagrams.
\subsection{THDM}
The phenomenological results for the
production processes
$\gamma \gamma \rightarrow \phi_i \phi_j$
with CP-even Higgses
$\phi_{i,j} \equiv h, H$
in the THDM are analyzed
in the following subsection.
\subsubsection{Production
cross-sections}
\begin{figure}[H]
\centering
\begin{tabular}{cc}
\\
\hspace{-3.5cm}
$\hat{\sigma}_{\phi_i \phi_j}(M^2=0)$ [pb]
&
\hspace{-3cm}
$\hat{\sigma}_{\phi_i \phi_j} (M^2=200^2)$
[pb]
\\
\includegraphics[width=8cm, height=6cm]
{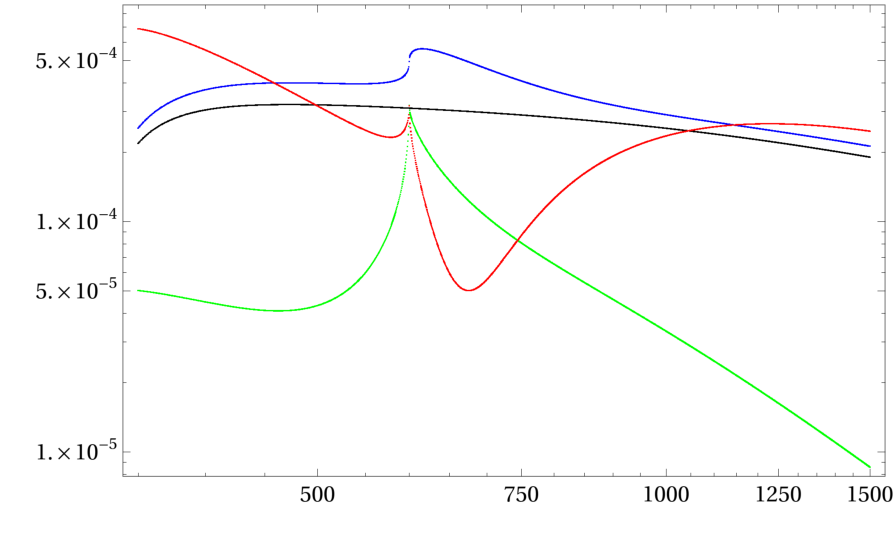}
&
\includegraphics[width=8cm, height=6cm]
{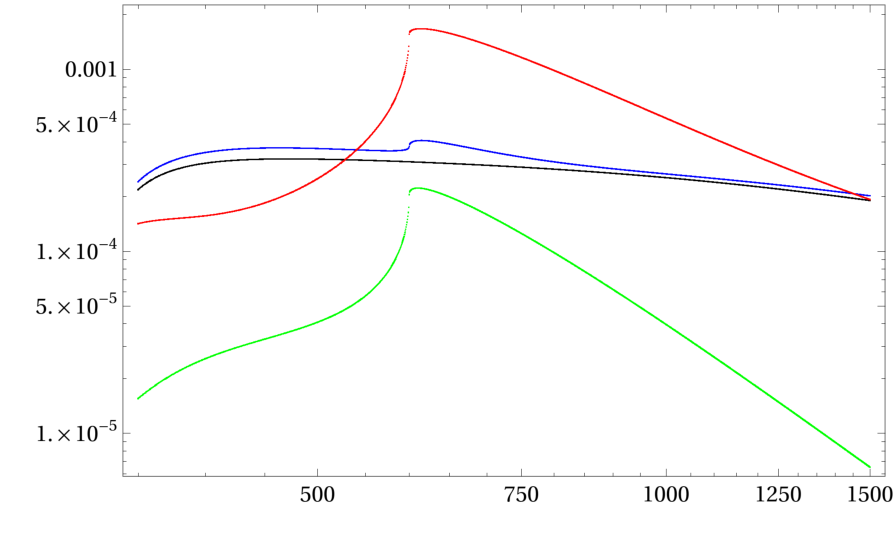}
\\
\hspace{6cm}
$\sqrt{\hat{s}_{\gamma\gamma}}$ [GeV]
&
\hspace{6cm}
$\sqrt{\hat{s}_{\gamma\gamma}}$
[GeV]
\\
&
\\
\hspace{-3cm}
$\hat{\sigma}_{\phi_i \phi_j}(M^2=500^2)$
[pb]
&
\hspace{-3.5cm}
$\hat{\sigma}_{\phi_i \phi_j} (M^2=700^2)$
[pb]
\\
\includegraphics[width=8cm, height=6cm]
{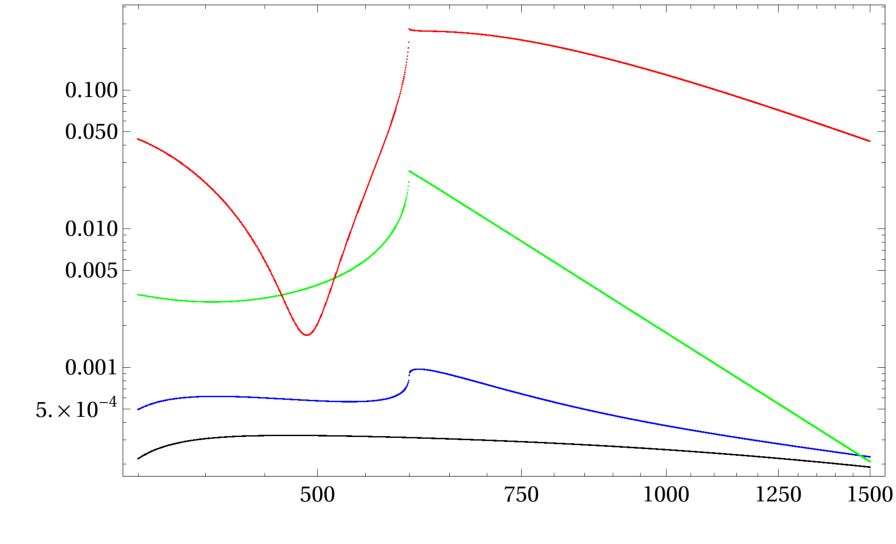}
&
\includegraphics[width=8cm, height=6cm]
{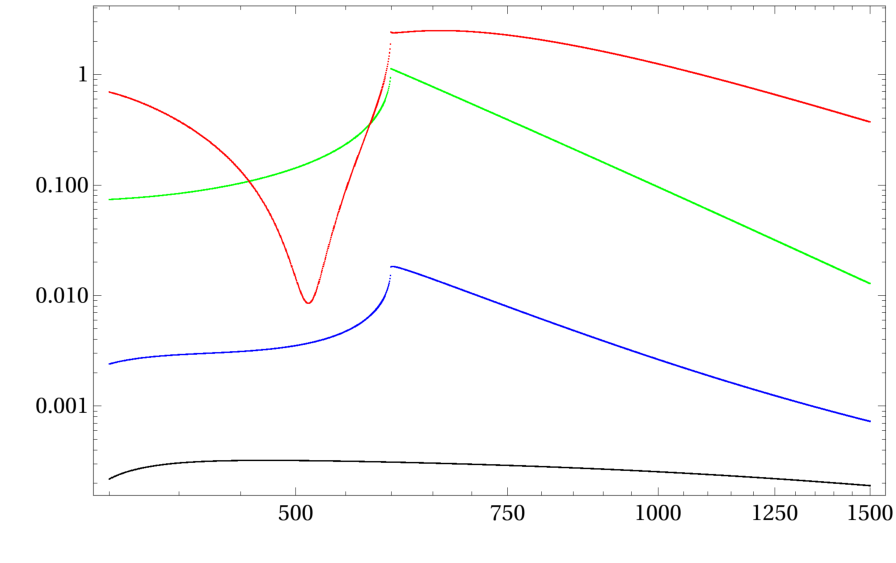}
\\
\hspace{6cm}
$\sqrt{\hat{s}_{\gamma\gamma}}$ [GeV]
&
\hspace{6cm}
$\sqrt{\hat{s}_{\gamma\gamma}}$ [GeV]
\end{tabular}
\caption{\label{totalcrosssections}
Cross-sections for $\gamma \gamma \rightarrow \phi_i\phi_j$ in the THDM, $\gamma \gamma \rightarrow hh$ in the SM are shown as functions of $\sqrt{\hat{s}_{\gamma\gamma}}$. In the plots, we vary $350$ GeV $\leq \sqrt{\hat{s}_{\gamma\gamma}} \leq 1500$ GeV. We select $M_{H^{\pm}}=300$ GeV, $t_{\beta}=5$ in this case. Moreover, we fix $M_H = 150$ GeV, $M^2 = 0, 200^2, 500^2, 700^2$ GeV$^2$ and take $c_{\beta-\alpha}=+0.2$ and $s_{\beta-\alpha} = +\sqrt{1- c_{\beta-\alpha}^2}$, accordingly.
}
\end{figure}

Cross-sections for $\gamma \gamma \rightarrow \phi_i \phi_j$ in the THDM are first investigated at several CoM energies. In Fig.~\ref{totalcrosssections}, $\hat{\sigma}_{\gamma \gamma \rightarrow \phi_i \phi_j}$ in the THDM together with $\hat{\sigma}_{\gamma \gamma \rightarrow hh}$ in the SM, are presented as functions of $\sqrt{\hat{s}_{\gamma\gamma}}$. The following data is generated at $M_{H^{\pm}}=300$ GeV, $M_H=150$ GeV and $t_{\beta}=5$. The CoM energies are varied as $350$ GeV $\leq \sqrt{\hat{s}_{\gamma\gamma}} \leq 1500$ GeV in the selected-configurations. The $Z_2$-breaking parameter $m_{12}^2 = M^2/s_{\beta}c_{\beta}$ is selected as follows: $M^2 = 0,~200^2,~500^2,~700^2$ GeV$^2$. In further, the mixing angle $\alpha$ is taken as $c_{\beta-\alpha}=+0.2$ and $s_{\beta-\alpha}=+\sqrt{1- c_{\beta-\alpha}^2}$, accordingly. The notations for all lines appearing in the presented plots are as follows: the black line shows the cross-sections of $\gamma \gamma \rightarrow hh$ in the SM, while the blue line presents $\gamma \gamma \rightarrow hh$ in the THDM. Additionally, the green (red) line presents $\gamma \gamma \rightarrow hH$ ($\gamma \gamma \rightarrow HH$) in the THDM, respectively. Generally, we observe that $\hat{\sigma}_{\phi_i \phi_j}$ are enhanced at $\sqrt{\hat{s}_{\gamma\gamma}} \sim 2 M_{H^{\pm}} = 600$ GeV for all cases. Among the productions, the data shows that cross-sections for $\gamma \gamma \rightarrow hH$ are suppressed compared with other productions, as a consequence of softly breaking the $Z_2$-symmetry. However, $\hat{\sigma}_{hH}$ becomes more and more significant once $M^2$ reaches large values.

We inspect the data in the case of $M^2=0$. One notices that $\hat{\sigma}_{HH}$ becomes largest in the regions $\sqrt{\hat{s}_{\gamma\gamma}} \leq \sim 450$ GeV and they decrease rapidly in the regions $\sqrt{\hat{s}_{\gamma\gamma}} \geq 450$ GeV. Moreover, $\hat{\sigma}_{hh}$ in the SM and the THDM are dominant in the regions of $\sqrt{\hat{s}_{\gamma\gamma}} \geq 450$ GeV contrasted to the ones for $\gamma \gamma \rightarrow hH,~HH$ in the THDM. Among the mentioned cross-sections, the $hh$ production in the THDM is largest in this case.

When $M^2=200^2$ GeV$^2$, the cross-sections for $HH$ productions in THDM become largest in comparison with other ones. These massive contributions are attributed to charged Higgs in the loop. Due to the $Z_2$-symmetry, the productions of $hH$ in the THDM are still suppressed in this case. For high regions of $M^2$, taking $M^2=500^2,~700^2$ GeV$^2$ as examples, the productions $\gamma \gamma \rightarrow hH,~HH$ become more and more dominant in comparison with $hh$ production in the SM and in the THDM.
\subsubsection{Enhancement factors} 
\begin{figure}[H]
\centering
\begin{tabular}{ccc}
\hspace{-4cm}
$\mu_{hh}^{\textrm{THDM}}$
&
\hspace{-4cm}
$\mu_{hH}^{\textrm{THDM}}$
&
\hspace{-4cm}
$\mu_{HH}^{\textrm{THDM}}$
\\
\includegraphics[width=5.5cm, height=5.5cm]
{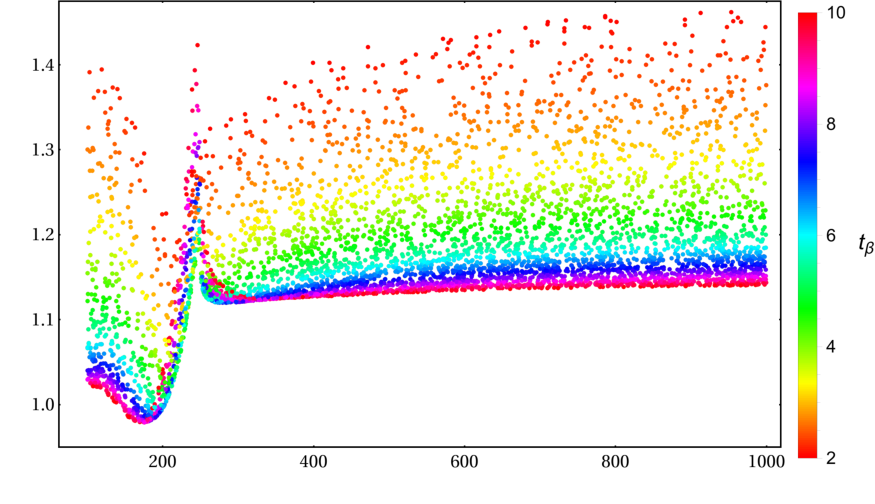}
&
\includegraphics[width=5.5cm, height=5.5cm]
{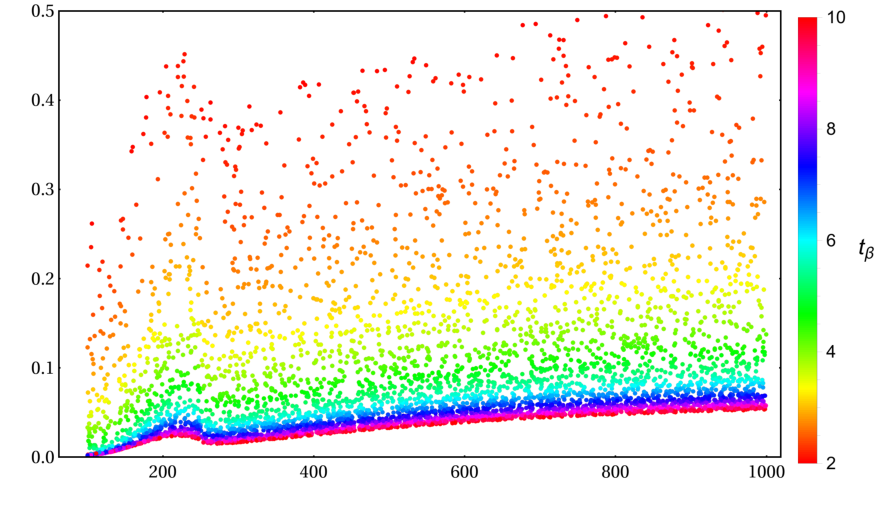}
&
\includegraphics[width=5.5cm, height=5.5cm]
{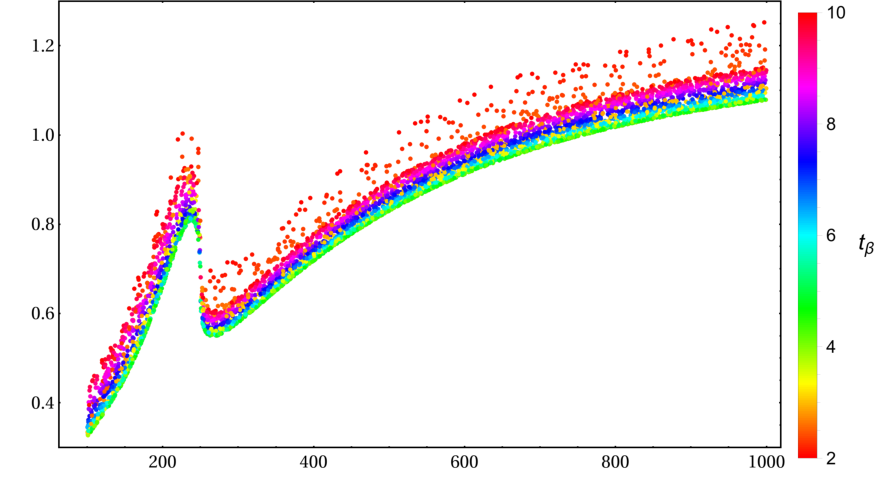}
\\
\hspace{3cm}
$M_{H^{\pm}}$ [GeV]
&
\hspace{3cm}
$M_{H^{\pm}}$
[GeV]
&
\hspace{3cm}
$M_{H^{\pm}}$
[GeV]
\\
&
\\
\hspace{-4cm}
$\mu_{hh}^{\textrm{THDM}}$
&
\hspace{-4cm}
$\mu_{hH}^{\textrm{THDM}}$
&
\hspace{-4cm}
$\mu_{HH}^{\textrm{THDM}}$
\\
\includegraphics[width=5.5cm, height=5.5cm]
{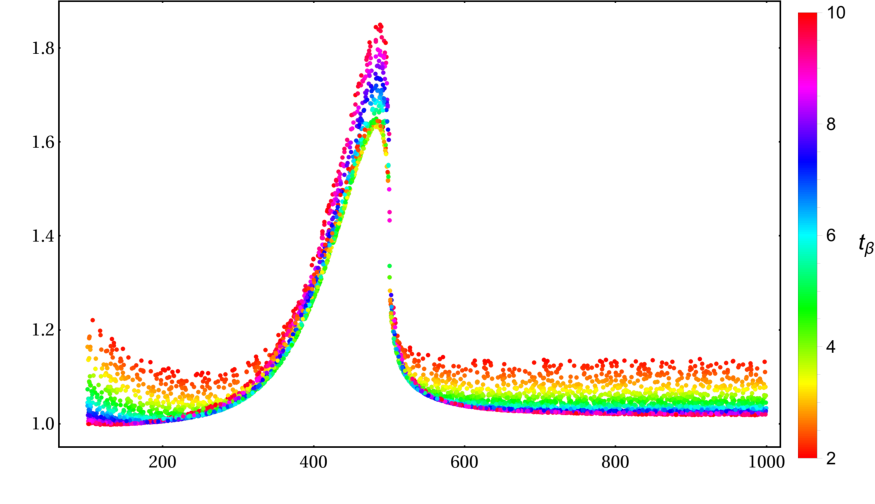}
&
\includegraphics[width=5.5cm, height=5.5cm]
{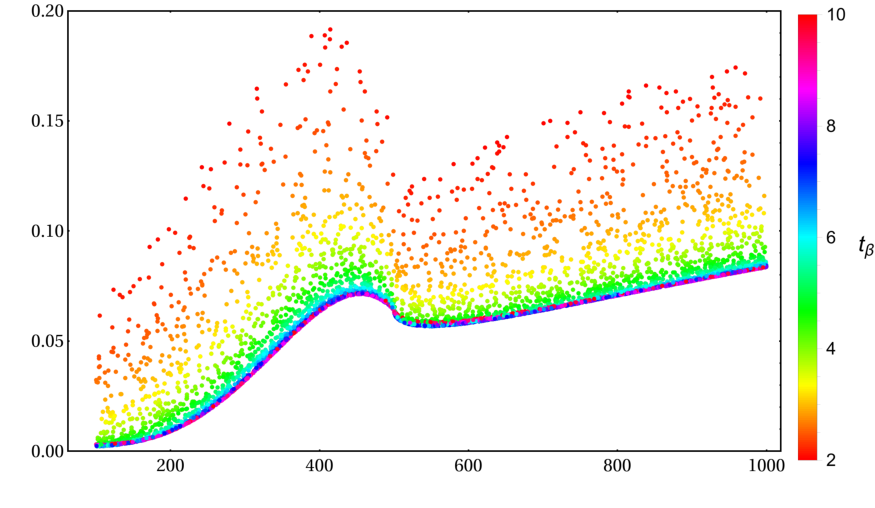}
&
\includegraphics[width=5.5cm, height=5.5cm]
{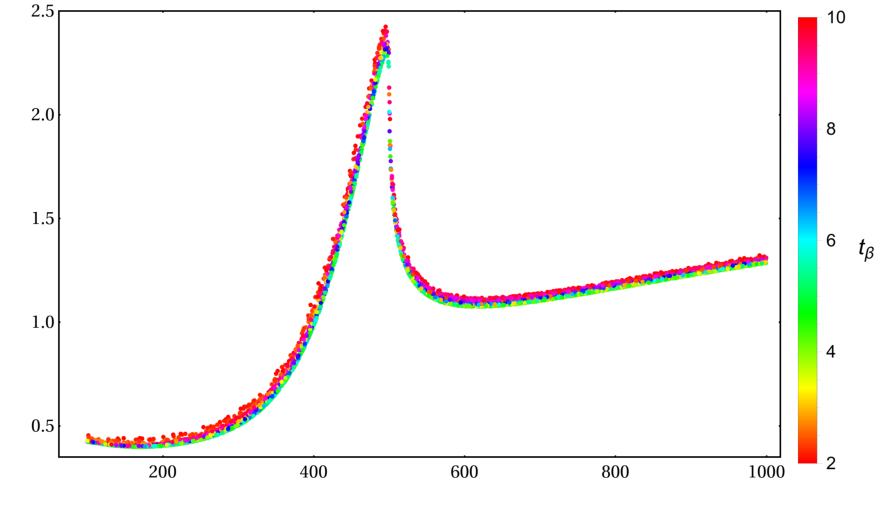}
\\
\hspace{3cm}
$M_{H^{\pm}}$ [GeV]
&
\hspace{3cm}
$M_{H^{\pm}}$
[GeV]
&
\hspace{3cm}
$M_{H^{\pm}}$
[GeV]
\end{tabular}
\caption{\label{figMHtbeta1cosplus}
The enhancement factors are presented in the parameter space of $M_{H^{\pm}},~t_{\beta}$. In the plots, we fix $M^2 = M_H^2 = 200^2$ GeV$^2$ and take $c_{\beta-\alpha} = +0.2$ and $s_{\beta-\alpha} = +\sqrt{1 - c_{\beta-\alpha}^2}$, accordingly. In these plots, we set $\sqrt{\hat{s}_{\gamma\gamma}} = 500$ GeV (for the above plots) and $\sqrt{\hat{s}_{\gamma\gamma}} = 1000$ GeV (for the below plots), respectively.
}
\end{figure}

\begin{figure}[H]
\centering
\begin{tabular}{ccc}
\hspace{-4.5cm}
$\mu_{hh}^{\textrm{THDM}}$
&
\hspace{-4.5cm}
$\mu_{hH}^{\textrm{THDM}}$
&
\hspace{-4.5cm}
$\mu_{HH}^{\textrm{THDM}}$
\\
\includegraphics[width=5.5cm, height=5.5cm]
{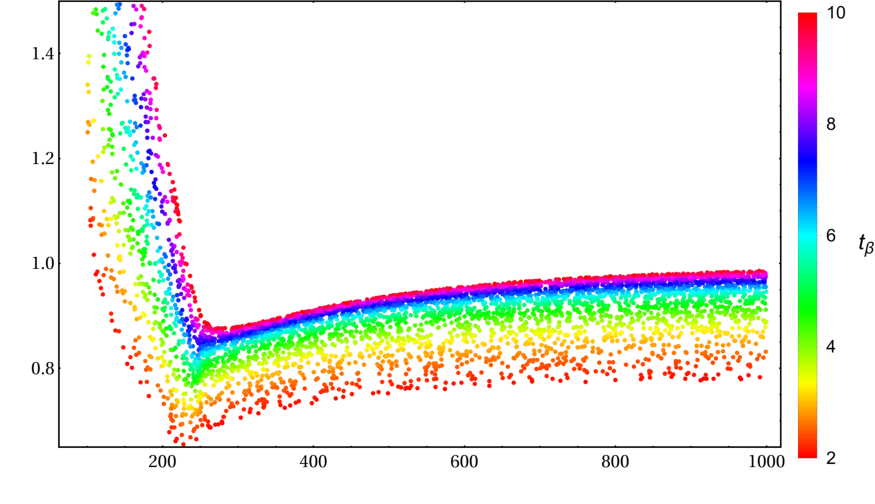}
&
\includegraphics[width=5.5cm, height=5.5cm]
{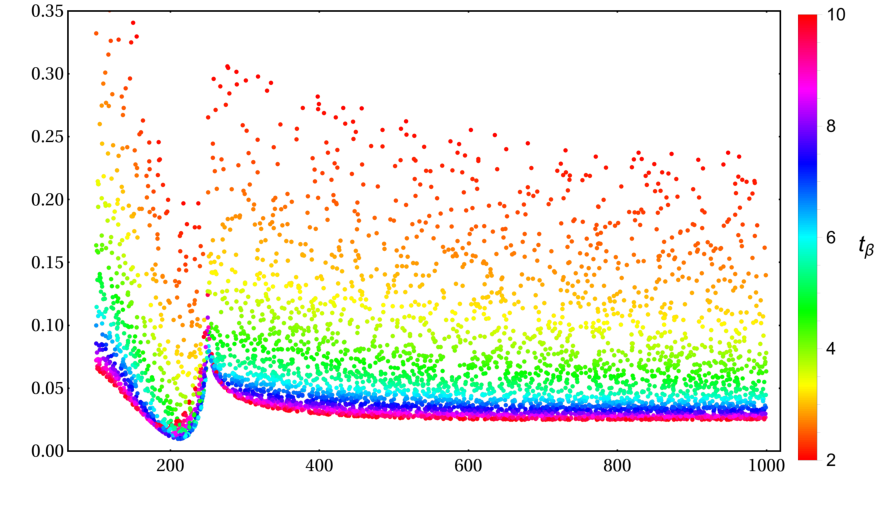}
&
\includegraphics[width=5.5cm, height=5.5cm]
{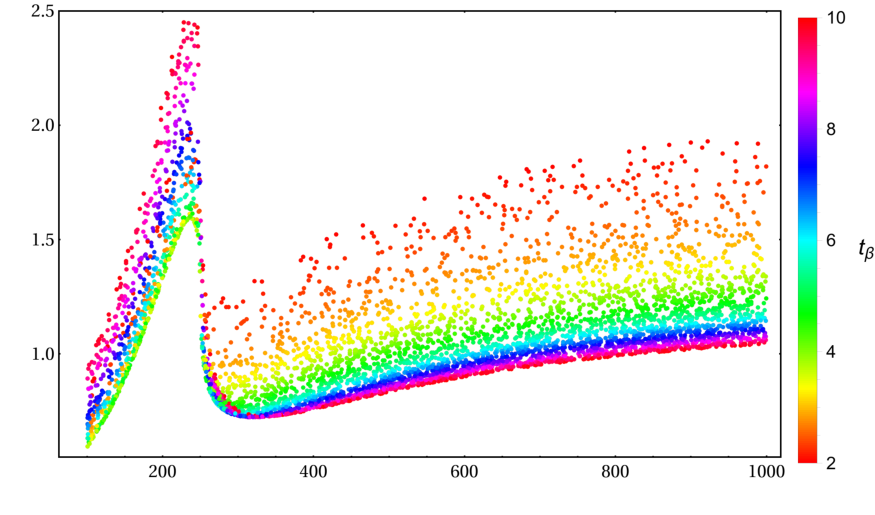}
\\
\hspace{3cm}
$M_{H^{\pm}}$ [GeV]
&
\hspace{3cm}
$M_{H^{\pm}}$
[GeV]
&
\hspace{3cm}
$M_{H^{\pm}}$
[GeV]
\\
\\
\hspace{-4.5cm}
$\mu_{hh}^{\textrm{THDM}}$
&
\hspace{-4.5cm}
$\mu_{hH}^{\textrm{THDM}}$
&
\hspace{-4.5cm}
$\mu_{HH}^{\textrm{THDM}}$
\\
\includegraphics[width=5.5cm, height=5.5cm]
{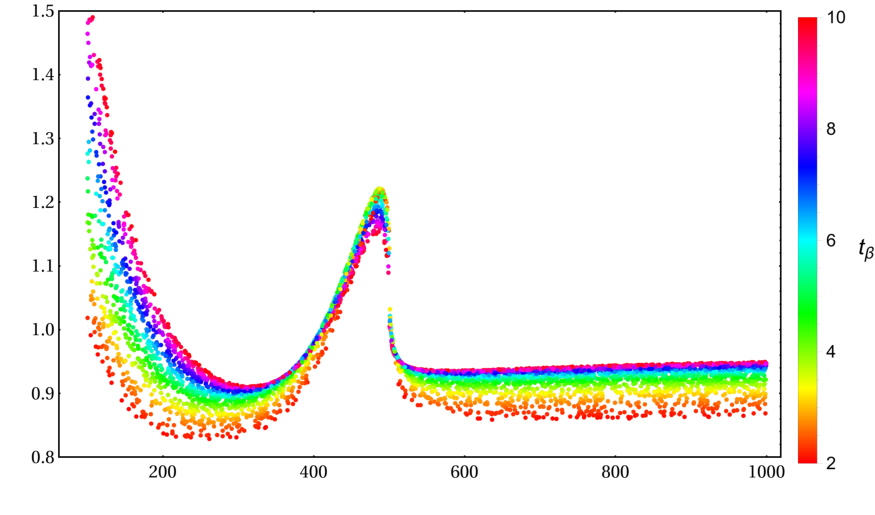}
&
\includegraphics[width=5.5cm, height=5.5cm]
{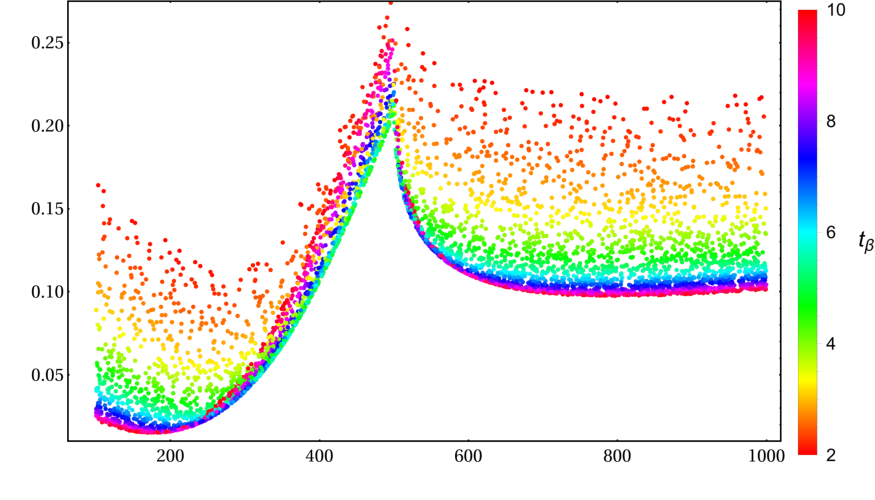}
&
\includegraphics[width=5.5cm, height=5.5cm]
{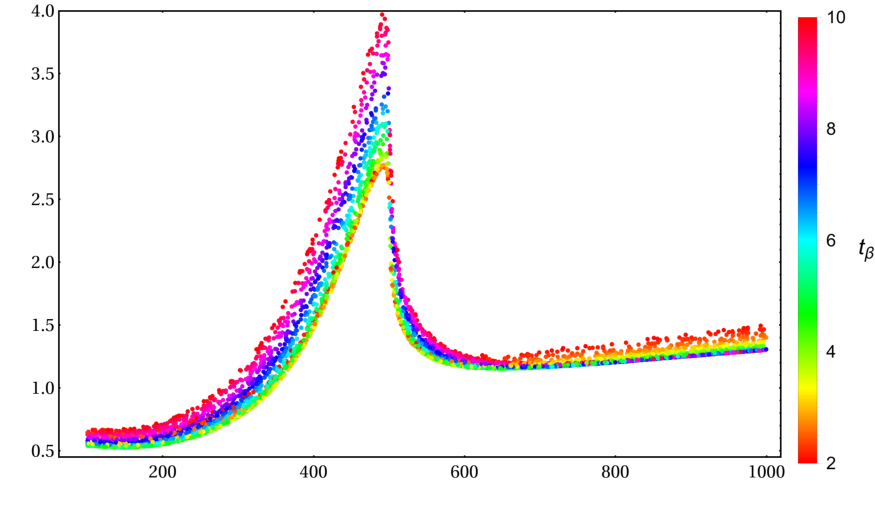}
\\
\hspace{3cm}
$M_{H^{\pm}}$ [GeV]
&
\hspace{3cm}
$M_{H^{\pm}}$
[GeV]
&
\hspace{3cm}
$M_{H^{\pm}}$
[GeV]
\end{tabular}
\caption{\label{figMHtbeta1cosmin}
The enhancement factors are presented in the parameter space $M_{H^{\pm}},~t_{\beta}$. In the plots, we consider the scenario for $c_{\beta-\alpha} = -0.2 < 0$ and $s_{\beta-\alpha} = +\sqrt{1 - c_{\beta-\alpha}^2}$, correspondingly. We also fix $M^2 = M_H^2 = 200^2$ GeV$^2$ and set $\sqrt{\hat{s}_{\gamma\gamma}} = 500$ (for all the above-plots) GeV and $\sqrt{\hat{s}_{\gamma\gamma}} = 1000$ GeV (for all the below-plots).
}
\end{figure}

We focus on investigating the enhancement factors defined in Eq.~\ref{EF_factors} for $\gamma \gamma \rightarrow \phi_i\phi_j$ in the THDM. The factors are first studied over the parameter space of $M_{H^{\pm}},~t_{\beta}$ in this subsection. Two scenarios for $c_{\beta-\alpha}>0$ and $c_{\beta-\alpha}<0$ are examined in detail. In Figs.~\ref{figMHtbeta1cosplus}, ~\ref{figMHtbeta1cosmin}, we fix $M^2=M_H^2=200^2$ GeV$^2$. Moreover, we vary $100$ GeV $\leq M_{H^{\pm}} \leq 1000$ GeV and set $2\leq t_{\beta}\leq 10$ in the following plots. The factors $\mu_{\phi_i\phi_j}^{\textrm{THDM}}$ are generated at $\sqrt{\hat{s}_{\gamma\gamma}} =500$ GeV (for all the above scatter-plots) and examined at $\sqrt{\hat{s}_{\gamma\gamma}} =1000$ GeV (for all the below scatter-plots). In the left panel, we show the enhancement factors for $hh$ production. In the middle (right) panel, the enhancement factors for $hH$ and $(HH)$
production are presented, respectively.

In Fig.~\ref{figMHtbeta1cosplus}, the first scenario for $c_{\beta-\alpha}>0$ is explored. In this scenario, we take $c_{\beta-\alpha} = +0.2$ as an example and $s_{\beta-\alpha} = +\sqrt{1 - c_{\beta-\alpha}^2}$, correspondingly. At $\sqrt{\hat{s}_{\gamma\gamma}} = 500$ GeV, $\mu_{hh}^{\textrm{THDM}}$ changes from $1$ to $1.5$ for the entire range of $M_{H^\pm}$. The values of $\mu_{hh}^{\textrm{THDM}}$ are enhanced around the peak at $M_{H^{\pm}} = \sqrt{\hat{s}_{\gamma\gamma}} / 2 = 250$ GeV. Predominantly, the factors are proportional to $t_{\beta}^{-1}$ in this case. Interestingly, we observe that the factors $\mu_{hH}^{\textrm{THDM}}$ change from $0$ to $0.5$ for the entire range of $M_{H^\pm}$. The suppressed values of $\mu_{hH}^{\textrm{THDM}}$ are expected, as explained in the previous paragraph, due to the $Z_2$-symmetry. The $\mu_{hH}^{\textrm{THDM}}$ exhibit the same behavior as $\mu_{hh}^{\textrm{THDM}}$, in that they are inversely proportional to $t_{\beta}$. On the other hand, the enhancement factors for $HH$ productions in the THDM are strongly dependent on the charged Higgs mass but change slightly with $t_{\beta}$. In the entire range of $M_{H^{\pm}}$, the factors $\mu_{HH}^{\textrm{THDM}}$ range from $0.3$ to $1.2$.

At $\sqrt{\hat{s}_{\gamma\gamma}} = 1000$ GeV, the factors $\mu_{hh}^{\textrm{THDM}}$ become the largest at the peak at $M_{H^{\pm}} = \sqrt{\hat{s}_{\gamma\gamma}} / 2 = 500$ GeV. Around the peak, $\mu_{hh}^{\textrm{THDM}}$ changes from $1.2$ to $1.8$. Beyond the peak, the factors change from $1.0$ to $1.2$ in the entire range of $M_{H^{\pm}}$. It is important to note that $\mu_{hh}^{\textrm{THDM}}$ slightly changes with $t_{\beta}$. For $HH$ productions, $\mu_{HH}^{\textrm{THDM}}$ varies from $0.4$ to $2.5$ around the peak (at $M_{H^{\pm}} = 500$ GeV) regions. It is noted that $\mu_{HH}^{\textrm{THDM}}$ slightly changes with $t_{\beta}$. Otherwise, $\mu_{hH}^{\textrm{THDM}}$ is
much smaller than $1$ and is inversely
proportional to $t_{\beta}$.
Another scenario for $c_{\beta-\alpha} < 0$ is considered for examining how the factors are affected by setting a different sign of $c_{\beta-\alpha}$ in this work. In Fig.~\ref{figMHtbeta1cosmin}, we take $c_{\beta-\alpha} = -0.2$ for an example and $s_{\beta-\alpha} = +\sqrt{1 - c_{\beta-\alpha}^2}$, accordingly. At $\sqrt{\hat{s}_{\gamma\gamma}} = 500$ GeV, it is interesting to observe that $\mu_{hh}^{\textrm{THDM}}$ shows different behavior in comparison with the previous scenario. At this CoM energy, the factors $\mu_{hh}^{\textrm{THDM}}$ can reach up to $1.5$ in the low region of $M_{H^{\pm}} < 200$ GeV. They then decrease to around $0.9$ when $M_{H^{\pm}} > 200$ GeV. There is no peak of the factors observed in this scenario because the contributions of singly charged Higgs exchanging in the one-loop triangle diagrams may cancel out with the ones from the one-loop box diagrams in this scenario. Surprisingly, we find that the factors $\mu_{hh}^{\textrm{THDM}}$ are proportional to $t_{\beta}$ in this scenario. For the $hH$ productions, the factors are suppressed and they are in the range of $[\sim 0.025, \sim 0.3]$. They are sensitive to $t_{\beta}^{-1}$ in all ranges
of charged Higgs mass. In $HH$ productions, it is found that the factors develop to the peak around $M_{H^{\pm}} = 500$ GeV. They reach a factor of $2.5$ around the peak and they are in the ranges of $[\sim 0.5, \sim 1.7]$ beyond the peak regions. In all
ranges of $M_{H^{\pm}}$, the factors
$\mu_{HH}^{\textrm{THDM}}$ are
proportional to $t_{\beta}^{-1}$ in this scenario.

The survey for all the enhancement factors at $\sqrt{\hat{s}_{\gamma\gamma}} = 1000$ GeV is presented
in the next paragraphs. The factors $\mu_{hh}^{\textrm{THDM}}$ are large in the regions ($M_{H^{\pm}} \leq 200$ GeV) and they can reach
up to $1.5$. They then decrease rapidly to the regions around $M_{H^{\pm}} \sim 300$ GeV and develop a peak at $M_{H^{\pm}} = \sqrt{\hat{s}_{\gamma\gamma}} / 2 = 500$ GeV. Around the peak, the enhancement factor is about $1.2$. In other ranges of charged Higgs mass, $\mu_{hh}^{\textrm{THDM}} \sim 0.9$. One also
observes that $\mu_{hh}^{\textrm{THDM}}$ is inversely
proportional to $t_{\beta}$ in this scenario. In $hH$
productions, the factors increase to the peak at $M_{H^{\pm}} = \sqrt{\hat{s}_{\gamma\gamma}} / 2 = 500$ GeV and they are about $0.3$ around the peak. In all regions of $M_{H^{\pm}}$, the mentioned factors are in the ranges of $[\sim 0.025, \sim 0.3]$ and they are inversely proportional to $t_{\beta}$. In the last case, it is found that the factors $\mu_{HH}^{\textrm{THDM}}$ show the same behavior
as the previous scenario. They are in the ranges of $[\sim 0.5, \sim 4]$ in all regions of $M_{H^{\pm}}$. However, the factors $\mu_{HH}^{\textrm{THDM}}$ depend only slightly on $t_{\beta}$ in this scenario.

\begin{figure}[H]
\centering
\begin{tabular}{ccc}
\hspace{-4.5cm}
$\mu_{hh}^{\textrm{THDM}}$
&
\hspace{-4.5cm}
$\mu_{hH}^{\textrm{THDM}}$
&
\hspace{-4.5cm}
$\mu_{HH}^{\textrm{THDM}}$
\\
\includegraphics[width=5.5cm, height=5.5cm]
{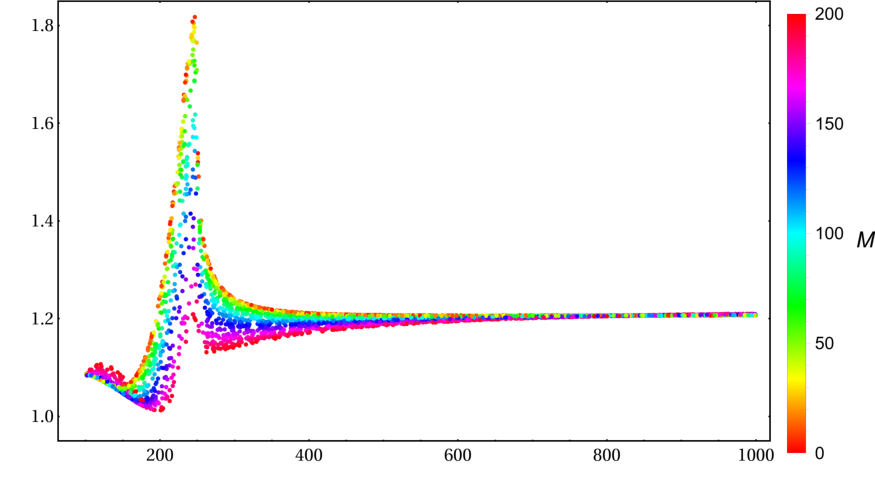}
&
\includegraphics[width=5.5cm, height=5.5cm]
{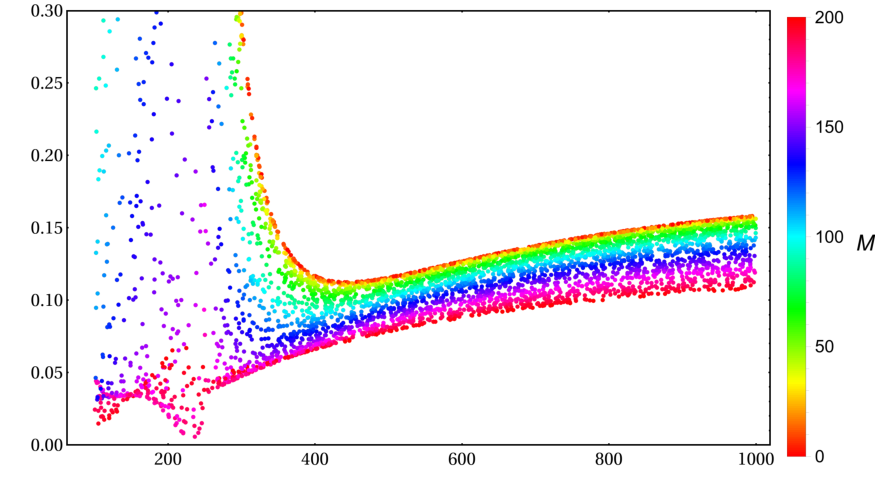}
&
\includegraphics[width=5.5cm, height=5.5cm]
{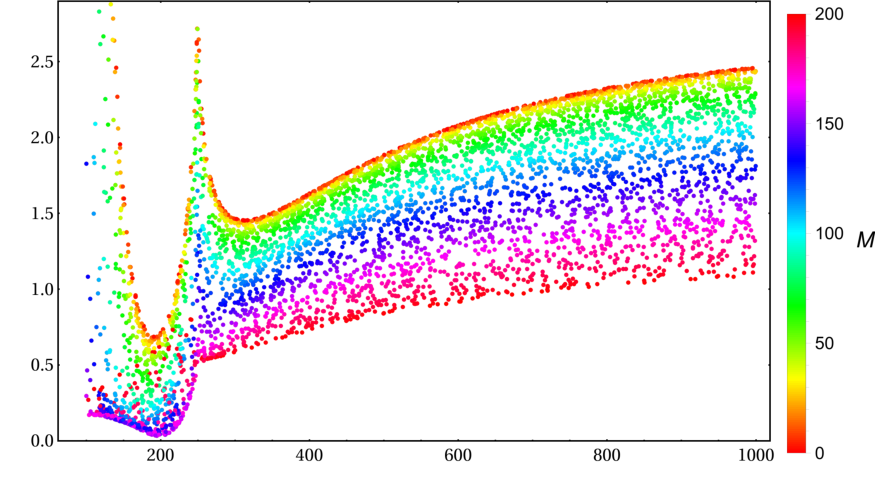}
\\
\hspace{3cm}
$M_{H^{\pm}}$ [GeV]
&
\hspace{3cm}
$M_{H^{\pm}}$
[GeV]
&
\hspace{3cm}
$M_{H^{\pm}}$
[GeV]
\\
\\
\hspace{-4.5cm}
$\mu_{hh}^{\textrm{THDM}}$
&
\hspace{-4.5cm}
$\mu_{hH}^{\textrm{THDM}}$
&
\hspace{-4.5cm}
$\mu_{HH}^{\textrm{THDM}}$
\\
\includegraphics[width=5.5cm, height=5.5cm]
{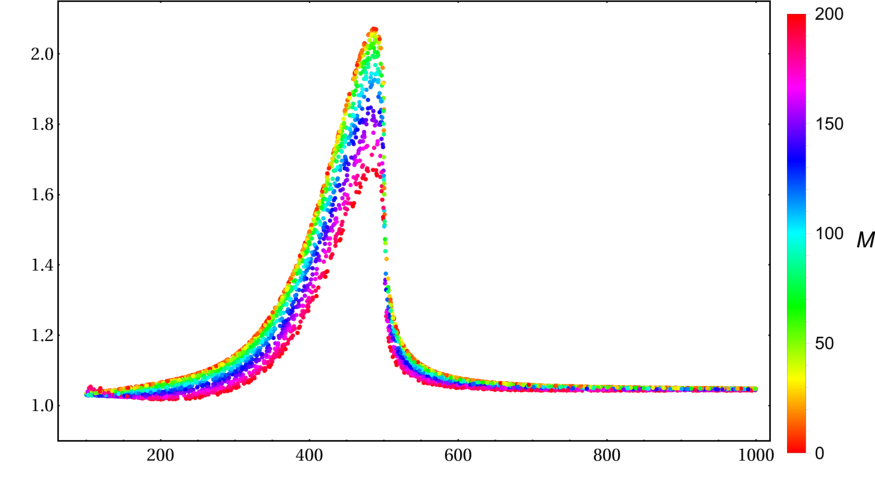}
&
\includegraphics[width=5.5cm, height=5.5cm]
{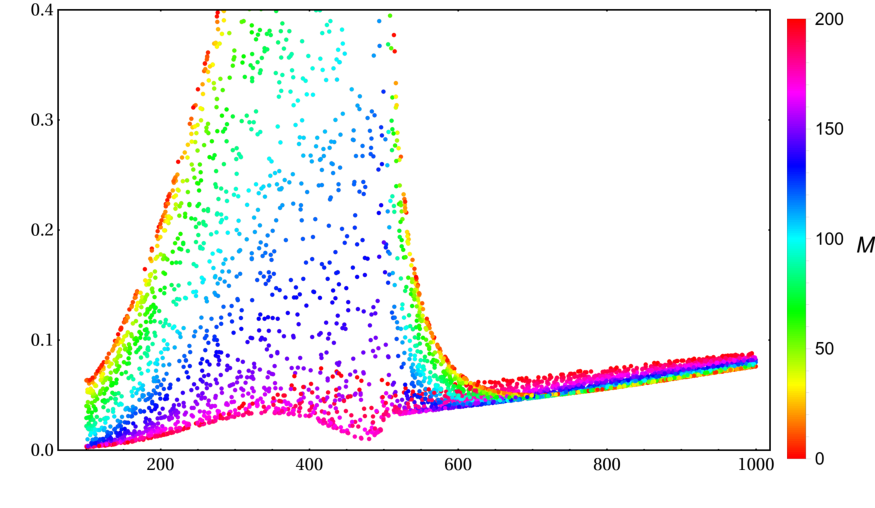}
&
\includegraphics[width=5.5cm, height=5.5cm]
{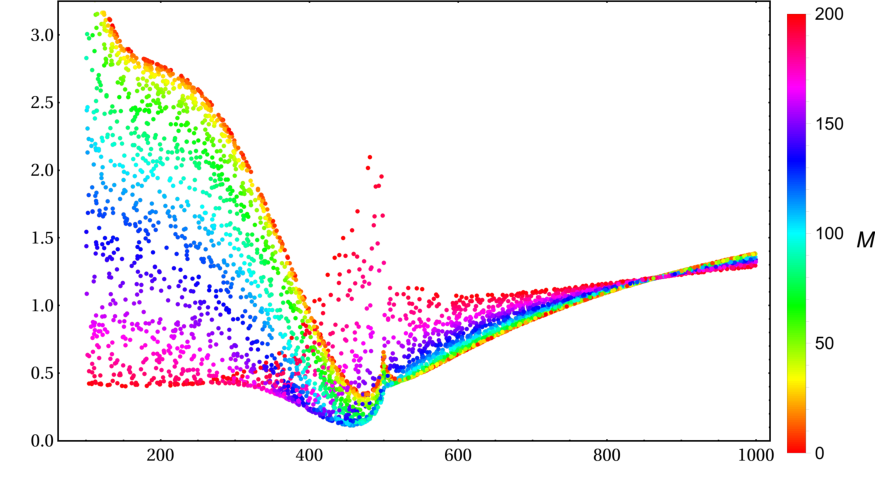}
\\
\hspace{3cm}
$M_{H^{\pm}}$ [GeV]
&
\hspace{3cm}
$M_{H^{\pm}}$
[GeV]
&
\hspace{3cm}
$M_{H^{\pm}}$
[GeV]
\end{tabular}
\caption{\label{figmucosplus}
The enhancement factors are presented
in the parameter space of
$M_{H^{\pm}},~M^2$.
In the plots, we take $t_{\beta}=5$ and
consider the first scenario
of $c_{\beta-\alpha}=+0.2$ and
$s_{\beta-\alpha}
= +\sqrt{1- c_{\beta-\alpha}^2}$,
accordingly.
In these plots, we set
$\sqrt{\hat{s}_{\gamma\gamma}} = 500$ GeV
(for the  above Figures) and
$\sqrt{\hat{s}_{\gamma\gamma}} = 1000$ GeV
(for the below Figures). }
\end{figure}

\begin{figure}[H]
\centering
\begin{tabular}{ccc}
\hspace{-4.5cm}
$\mu_{hh}^{\textrm{THDM}}$
&
\hspace{-4.5cm}
$\mu_{hH}^{\textrm{THDM}}$
&
\hspace{-4.5cm}
$\mu_{HH}^{\textrm{THDM}}$
\\
\includegraphics[width=5.5cm, height=5.5cm]
{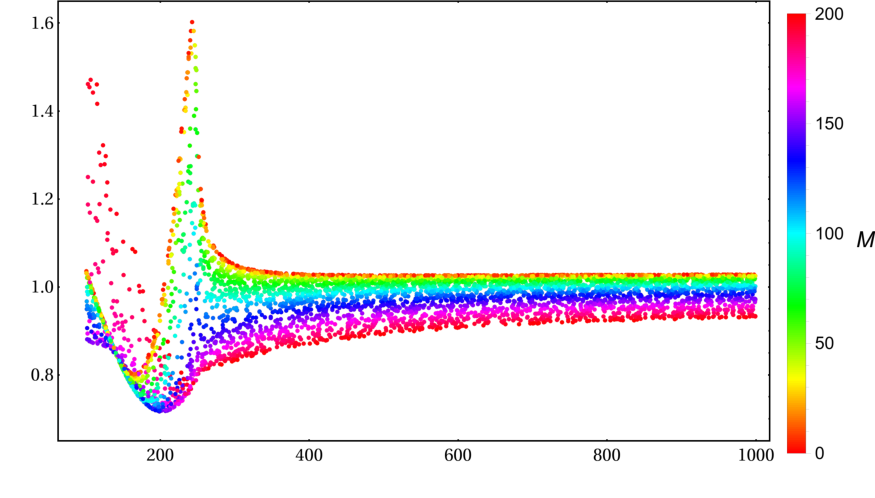}
&
\includegraphics[width=5.5cm, height=5.5cm]
{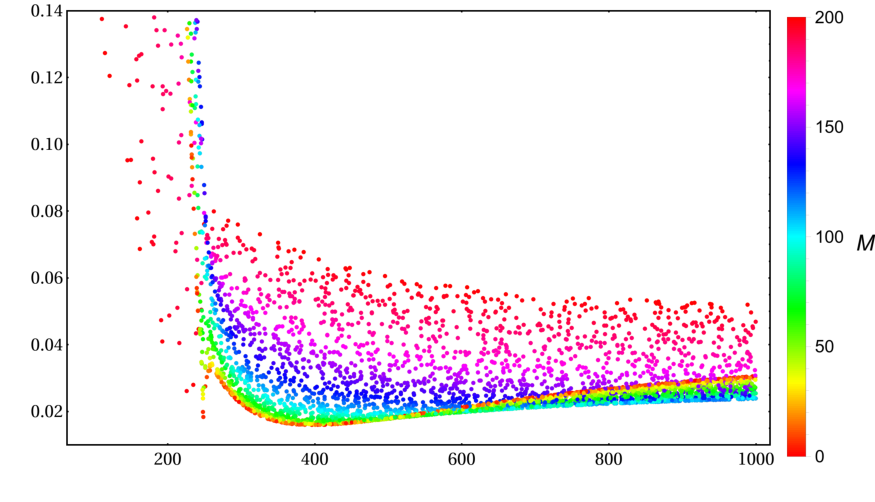}
&
\includegraphics[width=5.5cm, height=5.5cm]
{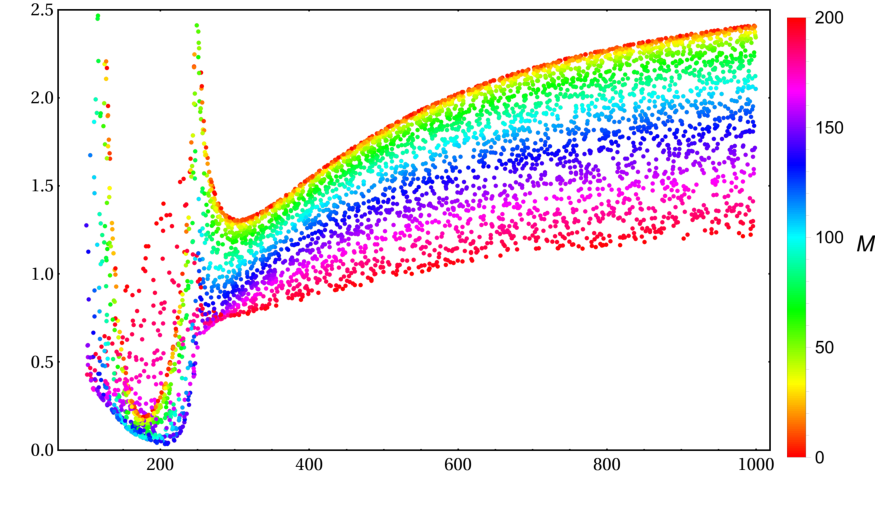}
\\
\hspace{3cm}
$M_{H^{\pm}}$ [GeV]
&
\hspace{3cm}
$M_{H^{\pm}}$
[GeV]
&
\hspace{3cm}
$M_{H^{\pm}}$
[GeV]
\\
\\
\hspace{-4.5cm}
$\mu_{hh}^{\textrm{THDM}}$
&
\hspace{-4.5cm}
$\mu_{hH}^{\textrm{THDM}}$
&
\hspace{-4.5cm}
$\mu_{HH}^{\textrm{THDM}}$
\\
\includegraphics[width=5.5cm, height=5.5cm]
{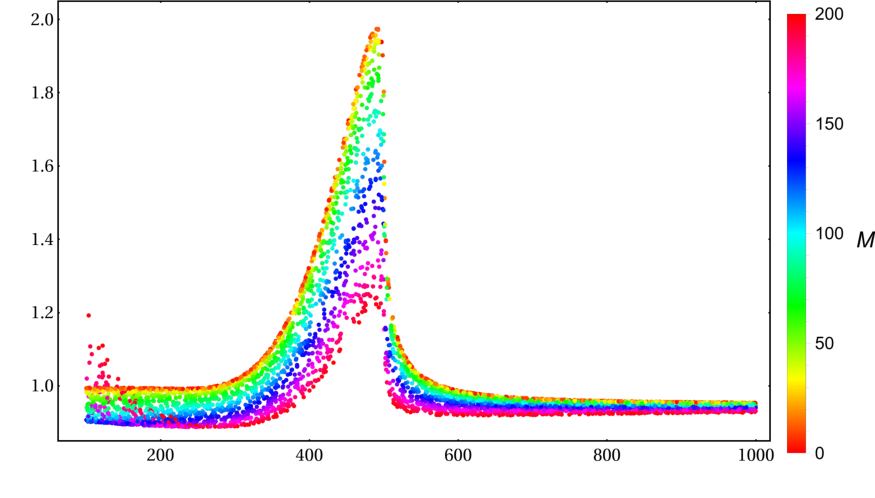}
&
\includegraphics[width=5.5cm, height=5.5cm]
{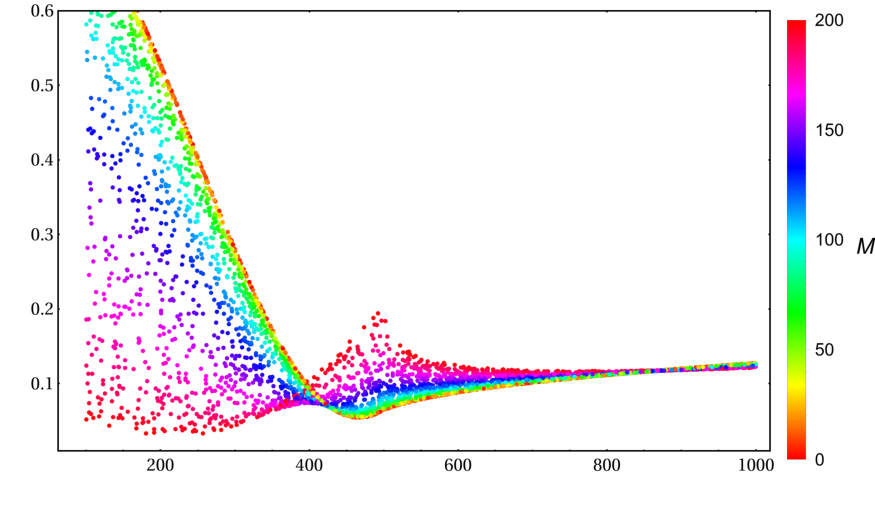}
&
\includegraphics[width=5.5cm, height=5.5cm]
{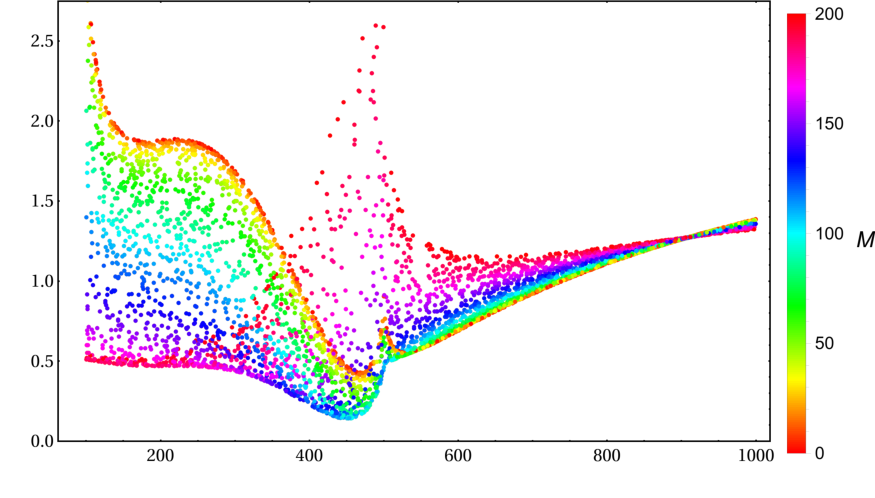}
\\
\hspace{3cm}
$M_{H^{\pm}}$ [GeV]
&
\hspace{3cm}
$M_{H^{\pm}}$
[GeV]
&
\hspace{3cm}
$M_{H^{\pm}}$
[GeV]
\end{tabular}
\caption{\label{figmucosminus}
The enhancement factors are presented
in the parameter space of $M_{H^{\pm}},~M^2$.
In the plots,
we take $t_{\beta}=5$ and
$c_{\beta-\alpha}=-0.2$
and $s_{\beta-\alpha}
=+\sqrt{1- c_{\beta-\alpha}^2}$,
accordingly.
In these plots, we set
$\sqrt{\hat{s}_{\gamma\gamma}}=500$ GeV
(the above Figures) and
$\sqrt{\hat{s}_{\gamma\gamma}} = 1000$ GeV
(the below Figures).}
\end{figure}

The enhancement factors scanned over the parameter space of $M_{H^{\pm}},~M^2$ in the THDM are also of great interest
in this work. Two scenarios for $c_{\beta-\alpha} > 0$ and $c_{\beta-\alpha} < 0$ are studied in detail in the following paragraphs. In Fig.~\ref{figmucosplus} (for the $c_{\beta-\alpha} > 0$ scenario) and Fig.~\ref{figmucosminus} (for the $c_{\beta-\alpha} < 0$ scenario), we consider $\sqrt{\hat{s}_{\gamma\gamma}} = 500$ GeV (for all the above scatter plots) and $\sqrt{\hat{s}_{\gamma\gamma}} = 1000$ GeV (for all the below scatter plots). Moreover, we vary the charged Higgs mass as $100$ GeV $\leq M_{H^{\pm}} \leq 1000$ GeV, the soft-breaking parameter as $0$ GeV$^2 \leq M^2 \leq 200^2$ GeV$^2$, and take $t_{\beta} = 5$ for all cases.

In Fig.~\ref{figmucosplus}, the first scenario of $c_{\beta-\alpha} > 0$ is examined. For this case, we take $c_{\beta-\alpha} = +0.2$ as an example and $s_{\beta-\alpha} = +\sqrt{1 - c_{\beta-\alpha}^2}$, accordingly. For $hh$ production at $\sqrt{\hat{s}_{\gamma\gamma}} = 500$ GeV, we observe the peak of $\mu_{hh}^{\textrm{THDM}}$ at $M_{H^{\pm}} = 250$ GeV, which is
corresponding to the threshold of cross-sections for $hh$ in the THDM at the peak $\sqrt{\hat{s}_{\gamma\gamma}} = 2M_{H^{\pm}}$. Around the peak, $\mu_{hh}^{\textrm{THDM}}$ varies from $1.0$ to $1.8$. Above the peak regions, the enhancement factors tend to $1.2$ and depend slightly on $M^2$. For $hH$ production, $\mu_{hH}^{\textrm{THDM}}$ is more sensitive to $M^2$ below the peak regions. The factors are in the range of $[0.07, 1.5]$ above
the peak regions. Around the peak, $\mu_{hH}^{\textrm{THDM}}$ can reach up to $0.3$. We also find the same behavior for $\mu_{HH}^{\textrm{THDM}}$. The factors for $HH$ production are large in the low regions of $M_{H^{\pm}}$ and around the peak $M_{H^{\pm}} = 250$ GeV. They are in the range of $[0.7, 2.4]$ above the peak regions. Generally, we observe that $\mu_{\phi_i\phi_j}^{\textrm{THDM}}$ is proportional to $1/M$ at this CoM energy.

At $\sqrt{\hat{s}_{\gamma\gamma}} = 1000$ GeV, we also find that $\mu_{hh}^{\textrm{THDM}}$ develops to the peak at $M_{H^{\pm}} = 500$ GeV, where the factors can reach up to $2.2$ and decrease rapidly beyond the peak. The factors depend slightly on $M^2$ and tend to $1$ beyond the peak regions. For $hH$ production, $\mu_{hH}^{\textrm{THDM}}$ is sensitive to $M^2$ in the peak regions.
They tend to $0.05$ and are slightly dependent on $M^2$ in the
regions above the peak. For $HH$ production, the factors become large in the regions below the peak and they are inversely proportional to $M^{-1}$. Around the peak, the factors are enhanced by large values of $M^2$. Above the peak regions, $\mu_{HH}^{\textrm{THDM}}$
varies around $1.0$.
Another scenario for $c_{\beta-\alpha} = - 0.2 < 0$ is also considered interestingly in this work. In Fig.~\ref{figmucosminus}, $s_{\beta-\alpha} = + \sqrt{1- c_{\beta-\alpha}^2}$ is obtained accordingly. We are going to comment on the physical results at $\sqrt{\hat{s}_{\gamma\gamma}} = 500$ GeV. For $hh$ production, we observe different behavior of $\mu_{hh}^{\textrm{THDM}}$ in comparison with the $c_{\beta-\alpha} > 0$ scenario. Specifically, the factors are large in the regions below the peak. Around the peak regions, they are enhanced by the small value of $M$ and can reach up to $1.6$. Above the peak regions, the factors are in the ranges of $[0.9, 1.1]$. We also observe the different behavior for the factors in $hH$ production compared with the previous scenario. The factors $\mu_{hH}^{\textrm{THDM}}$ reach large values in the below and around the peak regions and they are proportional to $M$. The factors $\mu_{hH}^{\textrm{THDM}}$ are in the ranges of $[0.2, 0.6]$ for $M_{H^\pm}$ in the above the peak regions. In the case of $HH$ production, $\mu_{HH}^{\textrm{THDM}}$ develops to the peak at $M_{H^\pm} \sim 250$ GeV. They are in the range of $[1.0,~2.4]$ in the above the peak regions. In general, the factors $\mu_{HH}^{\textrm{THDM}}$ depend slightly on charged Higgs mass and are proportional to $1/M$ in the above the peak regions.

We next comment on the physical results at $\sqrt{\hat{s}_{\gamma\gamma}} = 1000$ GeV. The factors $\mu_{hh}^{\textrm{THDM}}$ are enhanced by the small values of $M$ in the low regions of charged Higgs mass and they can reach up to $1.2$. They tend to $2.5$ around the peak. The factors are then varied around $\sim 0.9$. In general, the factors depend on $M^{-1}$. In the $hH$ production, $\mu_{hH}^{\textrm{THDM}}$ is more sensitive to $M^{-1}$ around the peak regions. They then tend to $0.2$ in the high mass regions of singly charged Higgs. For $HH$ production, the factors $\mu_{HH}^{\textrm{THDM}}$ strongly depend on $M^{-1}$. At the peak, the factors are enhanced by the large value of $M$. Above the peak regions, $\mu_{HH}^{\textrm{THDM}}$ tend to $\sim 1$.

\section{Conclusions} 
In this paper, we have presented the phenomenological results for one-loop induced processes $\gamma \gamma \rightarrow \phi_i\phi_j$ with CP-even Higgses $\phi_{i,j} \equiv h,~H$ at high-energy photon-photon collisions in the IHDM and the THDM. In the phenomenological results, we have shown the cross sections at several center-of-mass energies. The results show that cross sections for the computed processes in the models under investigation are enhanced around the threshold of charged Higgs pair production ($\sim 2 M_{H^\pm}$). Furthermore, the enhancement factors for the processes are examined in the parameter space of the models under consideration. In the IHDM, the factors are studied in the parameter space of $(M_{H^\pm},~\mu_2^2)$ and $(M_{H^\pm},~\lambda_2)$. In the THDM, the factors are analyzed
in the planes of $(M_{H^\pm},~t_{\beta})$ and $(M_{H^\pm},~M^2)$. Two scenarios of $c_{\beta-\alpha} > 0$ and $c_{\beta-\alpha} < 0$ have been studied in further detail. The factors give a different behavior when considering these scenarios. As a result, discriminations for the above-mentioned scenarios can be performed at future colliders.
The dependence of the cross section for the process $\gamma \gamma \rightarrow hH$ on $m_{12}^2$ provides a potential probe of the soft $Z_2$-breaking scale in the Two Higgs Doublet Model.
\\

\noindent
{\bf Acknowledgment:}~
This research is funded by Vietnam
National Foundation for Science and
Technology Development (NAFOSTED) under
the grant number $103.01$-$2023.16$.
\section*{Appendix A: Effective 
Lagrangian in the IHDM}         
The kinematic terms of Lagrangian in the IHDM
can be expanded as follows:
\begin{eqnarray}
\mathcal{L}_K^{\text{IHDM}}
&\supset&
\frac{2M_W^2}{v}
W^\pm_{\mu}W^{\mp,\mu}h
+
\frac{M_Z^2}{v}
hZ_{\mu}Z^{\mu}
+
i\frac{M_Zc_{2W}}{v}
Z^{\mu}
(H^{\mp}
\partial_{\mu}H^{\pm}
-
H^{\pm}
\partial_{\mu}H^\mp
)
\nonumber\\
&&
+
i\frac{M_Zs_{2W}}{v}
A^{\mu}
(H^{\mp}
\partial_{\mu}H^{\pm}
-
H^{\pm}\partial_{\mu}H^\mp
)
+
i
\frac{2M_Z^2c_W^2s_W}{v}
A^{\mu}W_{\mu}^{\pm}
G^{\pm}
\nonumber\\
&&
+
i
\frac{M_W}{v}
(HW^{\mp,\mu}\partial_{\mu}H^\pm
-H^{\pm}W^{\mp,\mu}\partial_{\mu}H
-HW^{\pm}_{\mu}\partial^{\mu}H^\mp
+H^{\mp}W^{\pm}_{\mu}\partial^{\mu}H)
\nonumber\\
&&
+\frac{M_Z^2s_{2W}^2}{v^2}
A_{\mu}A^{\mu}H^{\pm}H^{\mp}
+
\frac{2M_Z^2c_W^2s_W}{v^2}
HH^{\pm}
W_{\mu}^{\mp}A^{\mu}
+
\frac{2M_W^2}{v^2}
W^\pm_{\mu}W^{\mp, \mu}
hh
\nonumber\\
&&
+
\frac{2M_W^2}{v^2}
W^\pm_{\mu}W^{\mp, \mu}HH
+
\frac{M_Z^2s_{2W}^2}{v^2}
A_{\mu}A^{\mu}G^{\pm}G^{\mp}
\nonumber\\
&&
+
i\frac{M_Zs_{2W}}{v}A^{\mu}
(G^{\mp}\partial_{\mu}G^{\pm}
-G^{\pm}\partial_{\mu}G^{\mp})
+\cdots.
\end{eqnarray}
We also expand the scalar Higgs potential
of the IHDM
and collect the terms involving to
Higgs self-coupling as follows:
\begin{eqnarray}
-\mathcal{V}_{\text{IHDM}}
(\phi_1,\phi_2)
&\supset&
-\frac{3M_h^2}{v}hhh
+\frac{2(\mu_2^2-M_H^2)}{v}hHH
+\frac{2(\mu_2^2-M_{H^\pm}^2)}{v}
hH^{\pm}H^{\mp}
\nonumber\\
&&
+
\frac{M_{H^\pm}^2-M_H^2}{v}HH^\pm{G^\mp}
+\frac{2(\mu_2^2-M_{H^\pm}^2)}{v^2}
hhH^{\pm}H^{\mp}
-2\lambda_2HHH^{\pm}H^{\mp}
\nonumber\\
&&
-\frac{M_h^2}{v^2}hhG^{\pm}G^{\mp}
+\frac{2(\mu_2^2-M_{H^\pm}^2)}
{v^2}HHG^{\pm}G^{\mp}
+\cdots.
\end{eqnarray}
\section*{Appendix B: Effective Lagrangian  
in the THDM}                                
We expand the kinematic terms of Lagrangian  
in the THDM as follows:
\begin{eqnarray}
\mathcal{L}_{K}^{\text{THDM}}
&\supset&
\frac{2M_W^2}{v}
s_{\beta-\alpha}
hW_{\mu}^{\pm}W^{\mp,\mu}
+\frac{2M_W^2}{v}
c_{\alpha-\beta}
H W_{\mu}^{\pm}W^{\mp,\mu}
+
\frac{M_Z^2}{v}
s_{\beta-\alpha}
hZ_{\mu}Z^{\mu}
\nonumber\\
&&
+
\frac{M_Z^2}{v}c_{\alpha-\beta}
HZ_{\mu}Z^{\mu}
+
i \frac{M_Zc_{2W}}{v}
Z^{\mu}(H^{\mp}
\partial_\mu{H^\pm}-H^{\pm}
\partial_{\mu}H^{\mp})
\nonumber\\
&&
+i\frac{M_Zs_{2W}}{v}
A^{\mu}
(H^{\mp}\partial_\mu{H^\pm}
-H^{\pm}\partial_{\mu}H^{\mp})
+
\frac{4M_W^2s_W^2}{v^2}
H^{\pm}H^{\mp}A_\mu{A^{\mu}}
\nonumber\\
&&
-i
\frac{M_W s_{\beta-\alpha}}{v}
(HW^{\mp,\mu}\partial_{\mu}H^{\pm}
-HW^{\pm,\mu}\partial_{\mu}H^{\mp}
+H^{\mp}W^{\pm,\mu}\partial_{\mu}H
-H^{\pm}W^{\mp,\mu}\partial_{\mu}H)
\nonumber\\
&&
-i
\frac{M_Wc_{\beta-\alpha}}{v}
(-hW^{\mp,\mu}\partial_{\mu}H^{\pm}
+hW^{\pm,\mu}\partial_{\mu}H^{\mp}
-H^{\mp}W^{\pm,\mu}\partial_{\mu}h
+H^{\pm}W^{\mp,\mu}\partial_{\mu}h)
\nonumber\\
&&
+\frac{2M_W^2s_Wc_{\beta-\alpha}}{v^2}
hH^{\mp}W_{\mu}^{\pm}A^{\mu}
-
\frac{2M_W^2s_Ws_{\beta-\alpha}}{v^2}
HH^{\mp}W^{\pm}_{\mu}A^{\mu}
\nonumber\\
&&
+
\dfrac{2M_W^2}{v^2}W^\pm_{\mu}W^{\mp,\mu}HH
+\frac{2M_W^2}{v^2}W^\pm_{\mu}W^{\mp,\mu}hh
+ \cdots
\end{eqnarray}
Expanding 
the scalar potential in the THDM, 
we then collect the terms 
involving to the Higgs self-couplings
as 
\begin{eqnarray}
-\mathcal{V}_{\text{THDM}}(\phi_1,\phi_2)
&\supset&
-\lambda_{hHH}hHH
-\lambda_{Hhh}Hhh
-\lambda_{hH^{\pm}H^{\mp}}
hH^{\pm}H^{\mp}
\nonumber\\
&&
-\lambda_{hH^{\pm}H^{\mp}}
HH^{\pm}H^{\mp}
-\lambda_{HhH^{\pm}H^{\mp}}
HhH^{\pm}H^{\mp} + \cdots.
\end{eqnarray}
All coefficients of the mentioned 
couplings are shown explicitly 
in terms of physical parameters 
as follows:
\begin{eqnarray}
-\lambda_{hHH}
&=&
\frac{3\lambda_1v}{2}
s_{\alpha} c_{\alpha}^2 c_{\beta}
-\frac{3\lambda_2 v}{2}
s_{\beta}s_{\alpha}^2c_{\alpha}
-\frac{
\lambda_{345}
}{2}v
[c_{\beta}(2s_{\alpha}c_{\alpha}^2 -
s_{\alpha}^3) + s_{\beta}
(c_{\alpha}^3-2s_{\alpha}^2c_{\alpha})
]
\\
&&
\nonumber \\
&=&
\frac{s_{\alpha-\beta}
[s_{2\alpha}(3M^2-M_h^2-2M_H^2) +
M^2s_{2\beta}]}{v\;s_{2\beta}}
,
\\
-\lambda_{Hhh} &=&
-\frac{3\lambda_1v}{2}
c_{\beta}c_{\alpha}s_{\alpha}^2
-\frac{3\lambda_2v}{2}s_{\beta}
c_{\alpha}^2s_{\alpha}
-\frac{
\lambda_{345}
}{2}v
[s_{\beta}(s_{\alpha}^3
-2c_{\alpha}^2s_{\alpha})
-c_{\beta}(2c_{\alpha}s_{\alpha}^2
-c_{\alpha}^3)]
\\
&&
\nonumber \\
&=&
\frac{c_{\alpha-\beta}
[s_{2\alpha}(3M^2-M_H^2-2m_h^2)
-M^2s_{2\beta}]}{v\; s_{2\beta}}
,
\\
-\lambda_{HH^{\pm}H^{\mp}}
&=& -\lambda_1v
c_{\beta}c_{\alpha}s_{\beta}^2
-\lambda_2vs_{\beta}s_{\alpha}c_{\beta}^2
-\lambda_3v(s_{\beta}s_{\alpha}s_{\beta}^2
+c_{\beta}c_{\alpha}c_{\beta}^2)
+
\frac{\lambda_{45}}{2}
v s_{(2\beta)}\;
s_{\beta+\alpha}
\\
&=&
\frac{s_{\alpha+\beta}
(4M^2-3M_H^2-2M_{H^\pm}^2)
+(2M_{H^\pm}^2 - M_H^2)
s_{\alpha-3\beta}}
{2vs_{(2\beta)}},
\\
-\lambda_{hH^{\pm}H^{\mp}}
&=& \lambda_1v
c_{\beta}s_{\alpha}s_{\beta}^2
-\lambda_2 v s_{\beta}c_{\alpha}c_{\beta}^2
-\lambda_3v(s_{\beta}c_{\alpha}s_{\beta}^2
-c_{\beta}s_{\alpha}c_{\beta}^2)
+
\frac{\lambda_{45}
}{2}
v \; s_{(2\beta)}
\; c_{(\beta+\alpha)}
\\
&=&
\frac{c_{\alpha+\beta}
(4M^2-3M_h^2 - 2M_{H^\pm}^2)
+(2M_{H^\pm}^2-M_h^2)
c_{(\alpha-3\beta)}}
{2vs_{2\beta}},
\end{eqnarray}
and
\begin{eqnarray}
-\lambda_{HhH^{\pm}H^{\mp}}
&=&
\lambda_1s_{\beta}^2s_{\alpha}c_{\alpha}
-\lambda_2c_{\beta}^2s_{\alpha}c_{\alpha}
+\lambda_3s_{\alpha}c_{\alpha}c_{2\beta}
+(\lambda_4+\lambda_5)
s_{\beta}c_{\beta}c_{2\alpha}
\\
\label{gHhSS}
&=&
\frac{s_{2\alpha}(3c_{2\alpha}
+ c_{2(\alpha-2\beta)}
- 4c_{2\beta})}{4v^2s_{2\beta}^2} M_H^2
-\frac{s_{2\alpha}(3c_{2\alpha}
+ c_{2(\alpha-2\beta)}+4c_{2\beta})}
{4v^2s_{2\beta}^2} M_h^2
\nonumber\\
&&
+\frac{s_{2(\alpha-\beta)}}{v^2}
M_{H^\pm}^2 + \frac{(s_{2(\alpha-3\beta)}
+2s_{2(\alpha-\beta)}+5s_{2(\alpha+\beta)})}
{4v^2s_{2\beta}^2}M^2,
\\
-\lambda_{HhG^{\pm}G^{\mp}}
&=&
\lambda_1c_{\beta}^2s_{\alpha}c_{\alpha}
-\lambda_2s_{\beta}^2s_{\alpha}c_{\alpha}
-\lambda_3s_{\alpha}c_{\alpha}c_{2\beta}
-(\lambda_4+\lambda_5)s_{\beta}
c_{\beta}c_{2\alpha} \nonumber\\
\label{hHGG}
&=&\frac{1}{2v^2s_{2\beta}}
s_{2(\alpha-\beta)}
[(M_h^2- M_H^2)s_{2\alpha}
+ 2(M^2-M_{H^\pm}^2)s_{2\beta}].
\end{eqnarray}
Furthermore, we have the following
couplings:
\begin{eqnarray}
-\lambda_{hhh}
&=&
3v\Big[
\lambda_1s_{\alpha}^3c_{\beta}
-\lambda_2c_{\alpha}^3s_{\beta}
+(\lambda_3+\lambda_4+\lambda_5)
(s_{\alpha}c_{\beta}c_{\alpha}^2
-c_{\alpha}s_{\beta}s_{\alpha}^2)
\Big]
\nonumber\\
&=&
\dfrac{3e}{4M_Ws_Ws_{2\beta}}
\Big[
M^2c_{\alpha-3\beta}
+(M^2- M_h^2)c_{3\alpha-\beta}
+(2M^2-3 M_h^2)c_{\alpha+\beta}
\Big],
\end{eqnarray}
\begin{eqnarray}
-\lambda_{HHH}
&=&
-3v\Big[
\lambda_1c_{\alpha}^3c_{\beta}
+\lambda_2s_{\alpha}^3s_{\beta}
+(\lambda_3+\lambda_4+\lambda_5)
(c_{\alpha}c_{\beta}s_{\alpha}^2
+s_{\alpha}s_{\beta}c_{\alpha}^2)
\Big]
\nonumber\\
&=&
\dfrac{3e}{4M_Ws_Ws_{2\beta}}
\Big[ M^2s_{\alpha-3\beta}
+(M_H^2-M^2)s_{3\alpha-\beta}
+(2M^2-3M_H^2)s_{\alpha+\beta}
\Big],
\end{eqnarray}
\begin{eqnarray}
-\lambda_{HHH^\pm{H^\mp}}
&=&
-\lambda_1c_{\alpha}^2s_{\beta}^2
-\lambda_2s_{\alpha}^2c_{\beta}^2
-\lambda_3(c_{\alpha}^2c_{\beta}^2
+s_{\alpha}^2s_{\beta}^2)
+(\lambda_4+\lambda_5)
c_{\alpha}s_{\alpha}s_{2\beta} 
\nonumber\\
&=&
-\dfrac{2c_{\alpha-\beta}^2}{v^2}M_{H^\pm}^2
-\frac{s_{2\alpha}[3s_{2\alpha}
+s_{2(\alpha-2\beta)}-2s_{2\beta}]}
{4v^2s_{2\beta}^2} M_h^2 
\nonumber\\
&&-\dfrac{[c_{\alpha}^4
+c_{\alpha}^3s_{\alpha}\cot^3\beta
+c_{\alpha}s_{\alpha}^3\cot\beta
+s_{\alpha}^4\cot^4\beta]
\tan^2\beta}{v^2} M_H^2 
\nonumber\\
&&+\dfrac{s_{\beta}
[4c_{\alpha}c_{\beta}s_{\alpha}
+(1+\cot^4\beta)s_{\alpha}^2s_{\beta}
+c_{\alpha}^2(1+\cot^4\beta)
s_{\beta}\tan^2\beta]}{v^2}M^2,
\label{gHHHpHm}
\end{eqnarray}
\begin{eqnarray}
-\lambda_{hhH^{\pm}H^{\mp}}
&=&-\lambda_1s_{\alpha}^2s_{\beta}^2
-\lambda_2c_{\alpha}^2c_{\beta}^2
-\lambda_3(s_{\alpha}^2c_{\beta}^2
+c_{\alpha}^2s_{\beta}^2)
-(\lambda_4+\lambda_5)
c_{\alpha}s_{\alpha}s_{2\beta} \nonumber\\
&=&
-\dfrac{2s_{\alpha-\beta}^2}{v^2}M_{H^\pm}^2
-\dfrac{s_{2\alpha}
[3s_{2\alpha}
+ s_{2(\alpha-2\beta)}
+ 2s_{2\beta}]}
{4v^2s_{2\beta}^2} M_H^2
\nonumber\\
&&
+\dfrac{(-c_{\alpha}^4\cot^2{\beta}
+c_{\alpha}s_{\alpha}^3\cot\beta
+c_{\alpha}^3s_{\alpha}
\tan\beta-s_{\alpha}^4\tan^2\beta)}
{v^2}M_h^2
\nonumber\\
&&
+\dfrac{s_{\beta}c_{\beta}
[-4c_{\alpha}s_{\alpha}
+(1+\cot^4\beta)s_{\alpha}^2\tan^3{\beta}
+c_{\alpha}^2(\cot^3\beta+\tan\beta)]}
{v^2}M^2,
\label{ghhHpHm}
\end{eqnarray}
\begin{eqnarray}
-\lambda_{hhG^{\pm}G^\mp}
&=& -\lambda_1s_{\alpha}^2c_{\beta}^2
-\lambda_2c_{\alpha}^2s_{\beta}^2
-\lambda_3(c_{\alpha}^2c_{\beta}^2
+s_{\alpha}^2s_{\beta}^2)
+(\lambda_4+\lambda_5)
c_{\alpha}s_{\alpha}s_{2\beta}
\nonumber\\
&=&
\dfrac{2c_{\alpha-\beta}^2}
{v^2}(M^2 - M_{H^\pm}^2)
-\dfrac{c_{\alpha-\beta}^2
s_{2\alpha}}{v^2s_{2\beta}} M_H^2
+\dfrac{-3s_{2\beta} + 2s_{2\alpha}
+ s_{4\alpha-2\beta}}{4v^2s_{2\beta}}M_h^2,
\label{hhGpGm}
\end{eqnarray}
\begin{eqnarray}
-\lambda_{HHG^{\pm}G^\mp}
&=&-\lambda_1c_{\alpha}^2c_{\beta}^2
-\lambda_2s_{\alpha}^2s_{\beta}^2
-\lambda_3(s_{\alpha}^2c_{\beta}^2
+c_{\alpha}^2s_{\beta}^2)
-(\lambda_4+\lambda_5)
c_{\alpha}s_{\alpha}s_{2\beta}
\nonumber\\
&=&
\dfrac{2s_{\alpha-\beta}^2}{v^2}
(M^2- M_{H^\pm}^2)
+\dfrac{s_{\alpha-\beta}^2
s_{2\alpha}}{v^2s_{2\beta}}M_h^2
+\dfrac{-3s_{2\beta}-2s_{2\alpha}
+s_{4\alpha-2\beta}}
{4v^2s_{2\beta}}M_H^2.
\label{HHGpGm}
\end{eqnarray}
%
%
Additionally, we also derive 
the couplings relating to Goldstone 
bosons as follows:
\begin{eqnarray}
\mathcal{L}_K^{\text{THDM}}
&\supset&
\frac{2M_W^2s_W}{v}A_{\mu}W^{\pm, \mu}
G^{\mp}
+ \frac{4M_W^2s_W^2}{v^2}
A_{\mu}A^{\mu}G^\pm{G^\mp}
\nonumber\\
&&
+ i
\frac{2M_Ws_W}{v}
A^{\mu}(G^{\mp}
\partial_{\mu}G^\pm
- G^{\pm}
\partial_{\mu}G^{\mp})
+\cdots
\end{eqnarray}
From scalar potential, we have
\begin{eqnarray}
-\mathcal{V}(\phi_1,\phi_2)
&\supset&
- \lambda_{hH^{\pm}G^{\mp}}hH^{\pm}G^{\mp}
-\lambda_{HH^{\pm}G^{\mp}}HH^{\pm}G^{\mp}
+\cdots
\end{eqnarray}
where the coefficients of the couplings
are given by 
\begin{eqnarray}
-\lambda_{hH^{\pm}G^{\mp}}
&=&\frac{ec_{\alpha-\beta}}{2M_Ws_W}(M_{H^\pm}^2-M_h^2),
\\
-\lambda_{HH^{\pm}G^{\mp}}
&=&\frac{es_{\alpha-\beta}}{2M_Ws_W}(M_{H^\pm}^2-M_H^2).
\end{eqnarray}
\end{document}